\documentclass[twocolumn,showpacs,preprintnumbers,amsmath,amssymb]{revtex4}

\usepackage{graphicx}
\usepackage{dcolumn}
\usepackage{bm}

\begin{document}

\preprint{}
\title{Half-metallic ferromagnetism in binary compounds \\
of alkali metals with nitrogen: \textit{Ab initio} calculations}
\author{Krzysztof Zberecki, Leszek Adamowicz and Micha\l\ Wierzbicki}
\affiliation{Faculty of Physics, Warsaw University of Technology, ul. Koszykowa 75,
00-662 Warsaw, Poland}
\date{\today}

\begin{abstract}
The first-principles full-potential linearized augmented plane-wave method
based on density functional theory is used to investigate electronic
structure and magnetic properties of hypothetical binary compounds of I$^{A}$
subgroup elements with nitrogen (LiN, NaN, KN and RbN) in assumed three
types of cristalline structure (rock salt, wurtzite and zinc-blende). We
find that, due to the spin polarized \textit{p} orbitals of N, all four
compounds are half-metallic ferromagnets with wide energy bandgaps (up to
2.0 eV). The calculated total magnetic moment in all investigated compounds
for all three types of crystal structure is exactly 2.00 $\mu _{\text{B}}$
per formula unit. The predicted half-metallicity is robust
with respect to lattice-constant contraction. In all the cases ferromagnetic
phase is energetically favored with respect to the paramagnetic one. The
mechanism leading to half-metallic ferromagnetism and synthesis
possibilities are discussed.\newline
\end{abstract}

\pacs{71.15.Mb, 71.20.Dg, 72.25.Ba, 75.50.Cc}
\maketitle

\section{Introduction}

Half-metallic (HM) ferromagnets are materials in which, due to the
ferromagnetic decoupling, one of the spin subbands is metallic, whereas the
Fermi level falls into a gap of the other subband. The concept of HM
ferromagnet was first introduced by de Groot \textit{et al.} in 1983~\cite%
{degroot} on the basis of band structure calculations for NiMnSb and PtMnSb
semi-Heusler alloys. HM ferromagnets are considered as promising materials
to exploit the spin of charge carriers in new generations of transistors and
other integrated spintronic devices \cite{spintro}, in particular, as a
source of spin-polarized carriers injected into semiconductors, since only
one spin channel is active during charge transport, thus leading to 100\%\
spin-polarized electric current. 

Since 1983 many HM ferromagnets have been theoretically predicted, but very
few of them found experimental confirmation like metallic oxides CrO$_{2}$
 \cite{oxid1} and Fe$_{3}$O$_{4}$ \cite{oxid2} or manganese perovskite La$%
_{0.7}$Sr$_{0.3}$MnO$_{3}$ \cite{perov}. Many HM ferromagnetic materials
were predicted in transition-metals pnictides \cite{pni1,pni2} and
chalcogenides \cite{chalco1,chalco2} by means of first-principles
calculations. 

Recently, an unusual class of ferromagnetic materials \cite{hmf1,hmf2,hmf3,hmf4,hmf5,chang}, 
which do not contain transition-metal or rare-earth atoms, has been
predicted and analysed theoretically by \textit{ab initio} calculations. In
Ref. \cite{hmf1} the authors present \textit{ab initio} calculations for
CaAs in zinc-blende structure, where magnetic order is created with the main
contribution of the anion ${p}$ electrons (``${p}$-electron''
ferromagnetism). More comprehensive study was made in \cite{hmf2}, where the
authors investigate ${p}$-electron ferromagnetism in a number of
tetrahedrally coordinated binary compounds of I/II-V elements. The characteristic feature of this class of materials is the integer value of the total magnetic moment per formula unit, which, 
in some combinations of elements, can be as large as 3 $\mu_{\text{B}}$ \cite{chang}.  Results
presented in Ref. \cite{hmf3}, where magnetic and structural properties of II%
$^{A}$-V nitrides have been investigated using \textit{ab initio} methods,
motivated us to check pos\-si\-bi\-li\-ty of finding half-metallic ferromagnetism in
I$^{A}$-N binary compounds. Another motivation was that the atomic bonds in I%
$^{A}$-N nitrides are supposed to be mostly ionic in nature (due to
significant difference in electronegativity between I$^{A}$ atoms and
nitrogen atom) which was essential for appearing of ${p}$-electron
ferromagnetism in all previously considered HM cases (\cite{hmf1}-\cite{hmf5}%
). 

In this paper we present electronic structure and magnetic properties of
hypothetical I$^{A}$-N binary compounds (LiN, Nan, KN and RbN) with the rock
salt (RS), wurtzite (WZ) and zinc-blende (ZB) crystalline structure,
calculated by means of first-principles full-potential linearized augmented
plane-wave method. We find that all four compounds in all three types of
structures are HM ferromagnets with robust half-metallicity against lattice
compresion in the range from 2\% for LiN (RS) up to 50\% for NaN (WZ). This
is crucial parameter for the practice of epitaxial growth, \textit{e.g.} by
means of MBE or MOCVD, to fabricate magnetic ultrathin layer structures for
spin injection into suitable semiconductor substrate. 

The paper is organized as follows. Section II shows details of our
calculation method, while section III presents calculated total energy and
band structure. In section IV we analyse density of states and the origin of
half-metallic ferromagnetism. The paper is concluded with section V.

\section{Computational method}

All calculations were performed using WIEN2k code \cite{wien1} which
implements the full-potential linearized augmented plane wave (FLAPW) method 
\cite{flapw1}. Exchange and correlation were treated in the local spin
density approximation (LSDA) by adding gradient terms. This GGA
approximation was used in the Perdew-Burke-Ernzerhoff \cite{gga1}
parametrization. It should be mentioned that LSDA gives the magnetic moments
in a very good agreement with experiment but underestimates the lattice
constants in the case of transition metals. On the other hand, gradient
corrections significantly reduce this error and give the correct phase
stability but tend to overestimate the magnetic moment. We checked this
deficiency of GGA in our calculations and it turned out that it has a small
efect on the considered systems. 

The convergence of the basis set was controlled by the cut-off parameter $
RK_{\text{max}}$ = 8 together with 5000 k-point mesh for the integration over the
Brillouin zone. The angular momentum expansion up to $l$ = 10 and $G_{\text{max}}$
=12 a.u.$^{-1}$ ($G_{\text{max}}$=14 a.u.$^{-1}$ in case of RbN) for the potential
and charge density was employed in the calculations. Self-consistency was
considered to be achieved when the total energy difference between
succeeding iterations is less then 10$^{-5}$ Ry per formula unit. Geometry
optimization was performed allowing all atoms in the unit cell to relax,
constrained to the initially assumed crystal symmetry.

\section{Total energy and band structure}

For all four compounds investigated by us the total energies \textit{E} have
been calculated as a function of the lattice parameter \textit{a} for three
crystal structures, namely RS, WZ and ZB, in ferromagnetic and paramagnetic
state. We also optimized lattice constants \textit{a} by minimization of the
total energy \textit{E}. Figure 1 presents total energy \textit{E} versus
lattice constant \textit{a} for all three crystal structures. As one can see
in all compounds the RS strucure is the most energetically favored, thus,
indicating the ionic bonding as prevailed coupling mechanism. Also, in all
the cases, the paramagnetic phase is higher in energy than the corresponding
ferromagnetic one. Calculated equilibrium lattice constants \textit{a}$_{0}$
can be found in Table I as well as differences between total equilibrium
energies of ferromagnetic and paramagnetic states $\Delta E_{\text{tot}}^{%
\text{f-p}}$. 

The calculated total magnetic moment in all investigated compounds
for all three types of crystal structures is exactly 2.00 $\mu _{\text{B}}$
per formula unit. An integer value of magnetic moment is a characteristic
feature of HM ferromagnetism \cite{hmf2}. Table I shows total values of
magnetic moments for all compounds as well as contributions from I$^{A}$
atoms (Li, Na, K and Rb), N atom and interstitial region to total magnetic
moment. The main contribution in all cases comes from nitrogen anion ranging
from 1.51 $\mu _{\text{B}}$ for LiN in ZB to 1.85 $\mu _{\text{B}}$ for KN
in RS and ZB, which confirms the general feature that in HM ferromagnetic
compounds the most of resulting global magnetic moment is carried by anion
electrons. To verify that the
integer value of total magnetic moment stabilizes the crystal structure, we
have calculated the total energy \textit{E} as a function of the total
magnetic moment $\mu _{\text{tot}}$. Indeed, Figure 2 shows that the
creation of magnetic moment equal to 2.00 $\mu _{\text{B}}$ leads to the
minimization of the total energy in all the cases. 

>From the application point of view it is important to study
robustness of the half-metallicity with respect to lattice constant. Figure
3 shows magnetic moment as a function of lattice constant for all four
compounds in three crystal structures. In the case of LiN the total magnetic
moment remains integer until the lattice constant is compressed to critical
value of 2.75$\ {\rm\AA}$, 3.75$\ {\rm\AA}$ and
4.25$\ {\rm\AA}$ for WZ, ZB and RS structures, respectively. In
case of NaN, KN and RbN the values can be read from Figure 3. In all studied
compounds the most promising is wurtzite structure, where half-metallicity
is maintained up to contraction of the lattice parameter of 9\% (LiN), 54\%
(NaN), 22\% (KN) and 21\% (RbN). Wide bandgap semiconductors
AlN, GaN and ZnO in WZ crystal structure are considered as strong potential
materials for spintronic applications. The spin scattering relaxation time
of charge carriers during the transport process in these materials is
strongly increased as compared to GaAs \cite{relax}. 
This means longer lifetime of electrons in particular spin state
and longer spin ``memory" of electrons
when injected into these materials. Lattice constants of these wide bandgap semiconductors are
close to 3.2$\ {\rm\AA}$, what is in the range of LiN and NaN
HM ferromagnetism in WZ structure with 3\% and 18\% lattice mismatch,
respectively. Therefore, having in mind state-of-the-art in producing
artificial structures, at least some of  the HM ferromagnets studied here
seem to be potential candidates for epitaxial growth on AlN, GaN and ZnO
substrates as layers for injection of 100 \% spin polarized electrons. 
Most of ZB III-V semiconductors as well as narrow bandgap IV-VI semiconductors 
in RS structure (binary compounds of Pb with S, Se and Te) also satisfy lattice matching conditions as possible substrates 
for bulk growth of proposed here I$^{A}$-N compounds in HM ferromagnetic phase.

Figures 4 and 5 illustrate calculated spin resolved electron band structure for
RbN at equilibrium lattice constant for three types of crystal structures.
In all cases the minority spin subband is metallic, while the majority spin
subbands are separated from the Fermi level by wide bandgaps of 4.22 eV,
3.48 eV and 3.45 eV in RS, WZ and ZB crystal structure, respectively. This
is in contrast with the 3\textit{d} pnictides in which the bandgap appears
in the minority spin subband as a consequence of $d$ bands splitting \cite%
{pni2}. Three remaining compounds (LiN, NaN, and KN) also demonstrate
half-metallic nature with wide bandgaps. The HM bandgaps,  separating
allowed energies for majority spin states from the Fermi level, are wide
enough and ranging from about 0.2 eV for LiN (RS) to no less than 2.0 eV in
the case of KN and RbN for three considered crystal structures. The last
value is, to our knowledge, the largest one ever obtained for binary
compound HM ferromagnet, thus making KN and RbN very promising materials for
spintronic applications. The bands for RbN are flat and nearly dispersionless close to the Fermi level. The mechanism leading to this kind of energy bands has been discussed by other authors \cite{hmf2,hmf3} as an important condition for stability of HM ferromagnetism.

In the case of NaN and KN the bands look very
si\-mi\-lar to those presented in Figures 4 and 5 for RbN. For LiN the situation is
slightly different, because the bands near the Fermi level are more
dispersive, but far from breaking down HM ferromagnetism. 
Table II presents
calculated values of majority spin-up $E_{\text{g}}^{\text{up}}$ and
minority spin-down $E_{\text{g}}^{\text{dn}}$ subbands main bandgaps, as
well as HM bandgaps $E_{\text{g}}^{\text{HM}}$ in the majority-spin subband. 
Decrease of main bandgaps (of direct character only in the case of RS structure) and increase of HM bandgaps with increasing
atomic number of alkali metal elements is observed.

\begin{figure}[H]
\begin{center}
\begin{tabular}{cc}
\resizebox{80mm}{!}{\includegraphics[angle=270]{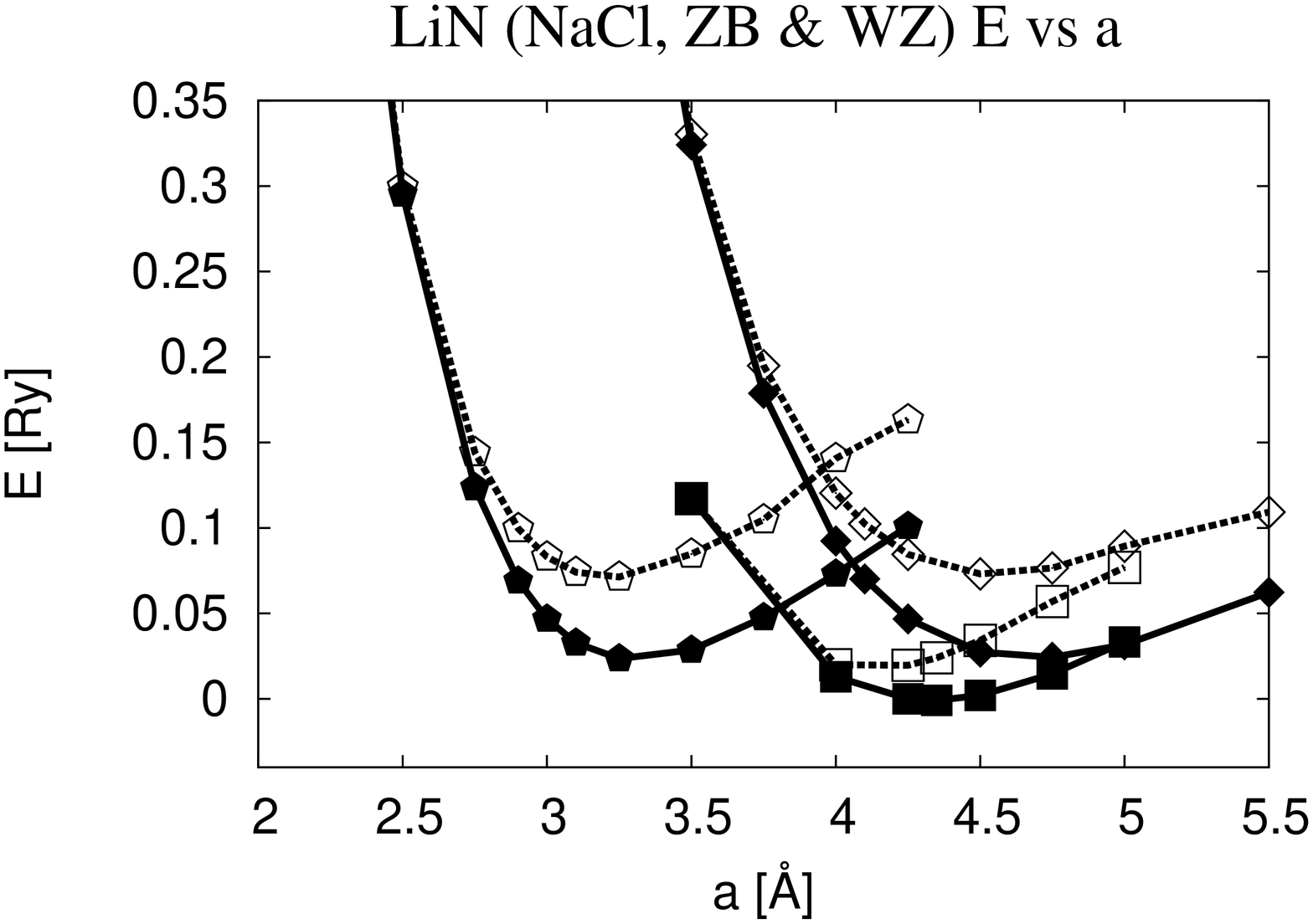}} 
\resizebox{80mm}{!}{\includegraphics[angle=270]{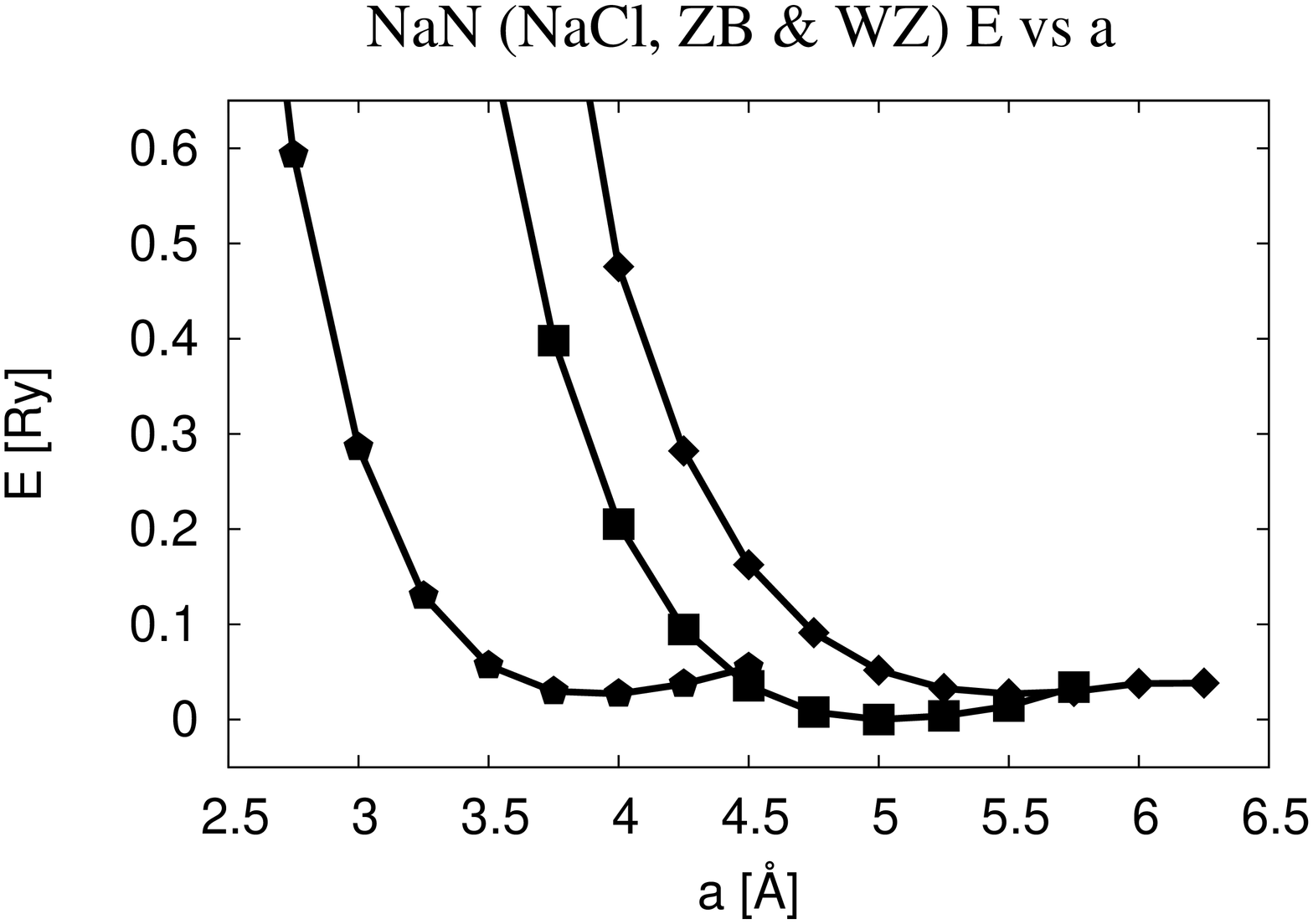}} &  \\ 
\resizebox{80mm}{!}{\includegraphics[angle=270]{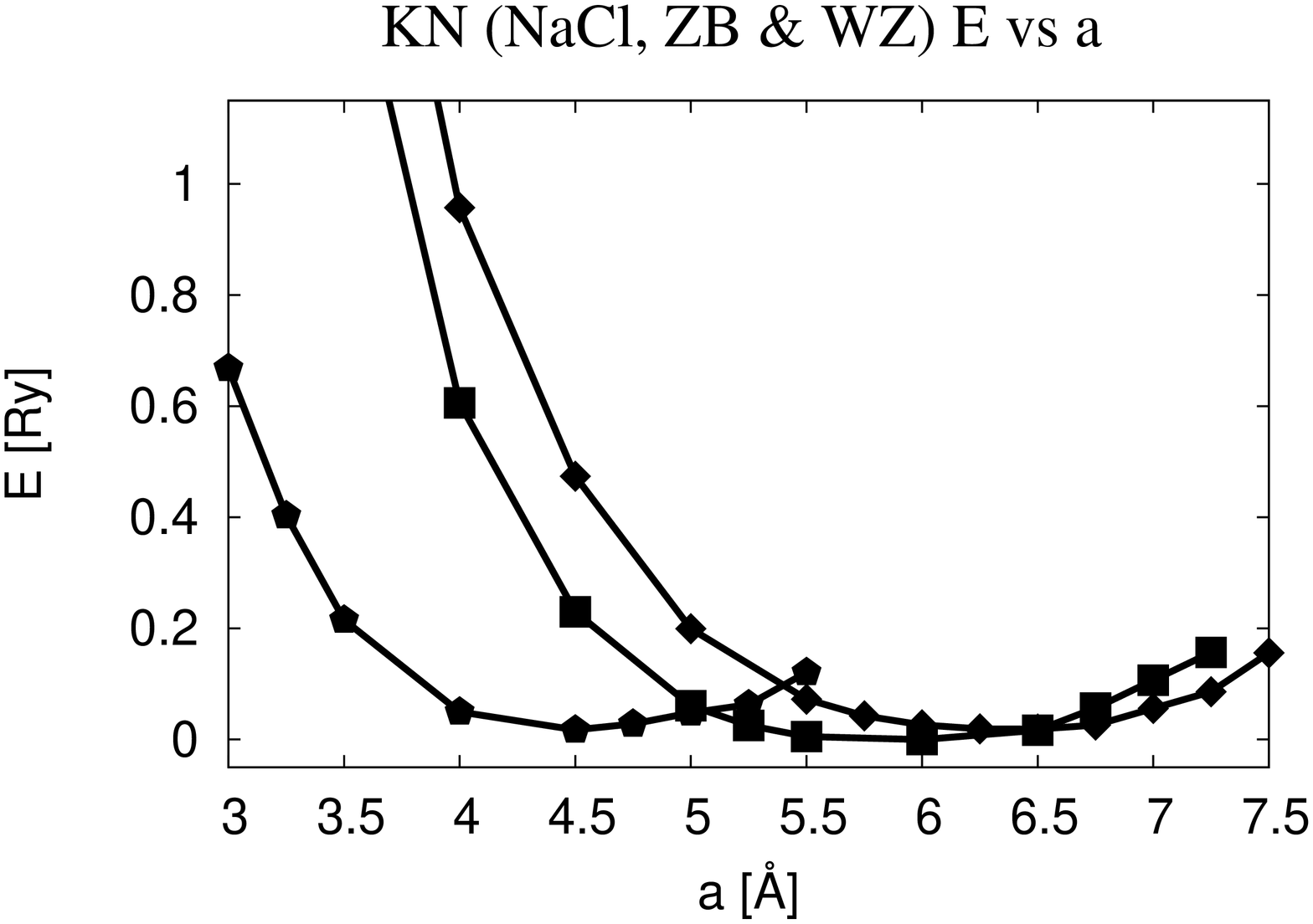}} 
\resizebox{80mm}{!}{\includegraphics[angle=270]{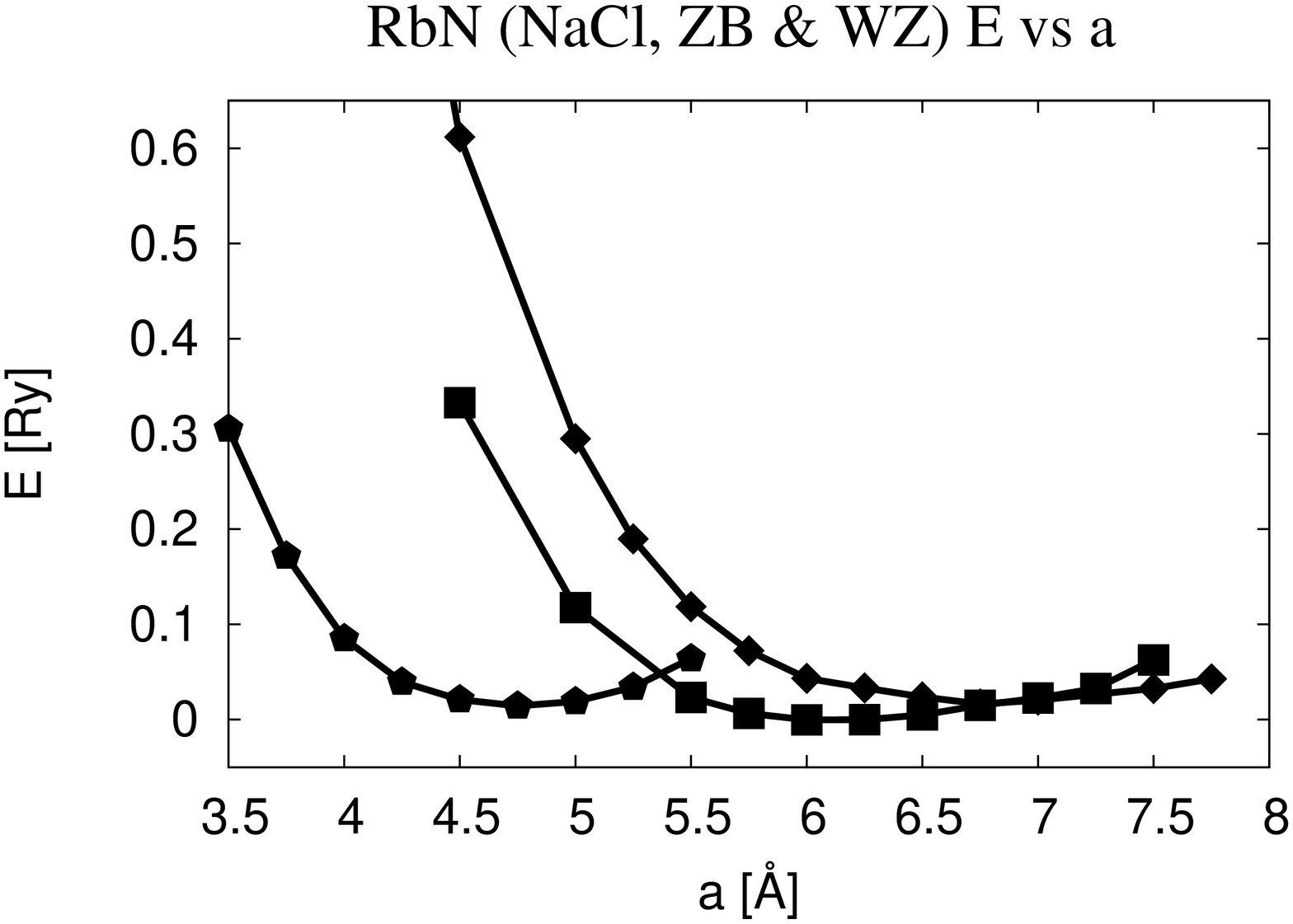}} &  \\ 
& 
\end{tabular}
\end{center}

\caption{Energy \textit{E} (in Ry) vs. lattice constant \textit{a} (in$\ 
{\rm \AA}$) for LiN, NaN, KN and RbN in three crystal structures
(squares -- RS, pentagons -- WZ, diamonds -- ZB, all for ferromagnetic phase;
open symbols -- the same, but for paramagnetic phase). Plots for paramagnetic
energies are similar, therefore we give them only in the case of LiN.}
\end{figure}

\begin{table}[H]
\begin{tabular}{lccccccc}
\hline
\hline
& $a_{0}$(${\rm \AA}$) & $\mu_{\text{tot}}(\mu_{\text{B}})$ & 
$\mu_{\text{I}^A}(\mu_{\text{B}})$ & \ $\mu_{N}(\mu_{\text{B}})$
&$\mu_{\text{int}}(\mu_{\text{B}})$ & 
$\Delta{E}_{\text{tot}}^{\text{f-p}}$(Ry) &  
\\ 
\hline
LiN &  &  &  &  &  &  &  \\ \hline
(RS) & 4.36 & 2.00 & 0.06 & 1.72 & 0.21 & $-0.025$ &  \\ 
(WZ) & 3.31 & 2.00 & 0.03  & 1.17 & 0.80 & $-0.102$ &  \\ 
(ZB) & 4.68 & 2.00 & 0.04 & 1.51 & 0.44 & $-0.052$ &  \\ \hline
NaN &  &  &  &  &  &  &  \\ \hline
(RS) & 5.03 & 2.00 & 0.01 & 1.73 & 0.26 & $-0.061$ &  \\ 
(WZ) & 3.91 & 2.00 & 0.02 & 1.47 & 0.51 & $-0.141$ &  \\ 
(ZB) & 5.49 & 2.00 & 0.01 & 1.67 & 0.31 & $-0.074$ &  \\ \hline
KN &  &  &  &  &  &  &  \\ \hline
(RS) & 5.82 & 2.00 & 0.02 & 1.85 & 0.12 & $-0.075$ &  \\ 
(WZ) & 4.51 & 2.00 & 0.03 & 1.75 & 0.22 & $-0.160$ &  \\ 
(ZB) & 6.38 & 2.00 & 0.03 & 1.85 & 0.11 & $-0.080$ &  \\ \hline
RbN &  &  &  &  &  &  &  \\ \hline
(RS) & 6.12 & 2.00 & 0.03 & 1.83 & 0.13 & $-0.074$ &  \\ 
(WZ) & 4.75 & 2.00 & 0.03 & 1.73 & 0.24 & $-0.156$ &  \\ 
(ZB) & 6.70 & 2.00 & 0.03 & 1.83 & 0.13 & $-0.077$ &  \\ \hline
\end{tabular}

\caption{Equilibrium lattice constants \textit{a}$_{0}$ (in$\ {\rm\AA}$) 
together with total $\protect\mu_{\text{tot}}$, partial
atomic-site resolved $\protect\mu_{\text{I}^{A}}$, $\protect\mu_{\text{N}}$
and interstitial $\protect\mu_{\text{int}}$ magnetic moments (in $\protect\mu
_{\text{B}}$) as well as total energy difference $\ \Delta{E}_{\text{tot}}^{
\text{f-p}}$ (Ry) between ferro- and paramagnetic state (in Ry) for RS, WZ
and ZB crystal structure.} 
\end{table}



\begin{figure}[H]
\begin{center}
\begin{tabular}{cc}
\resizebox{80mm}{!}{\includegraphics[angle=270]{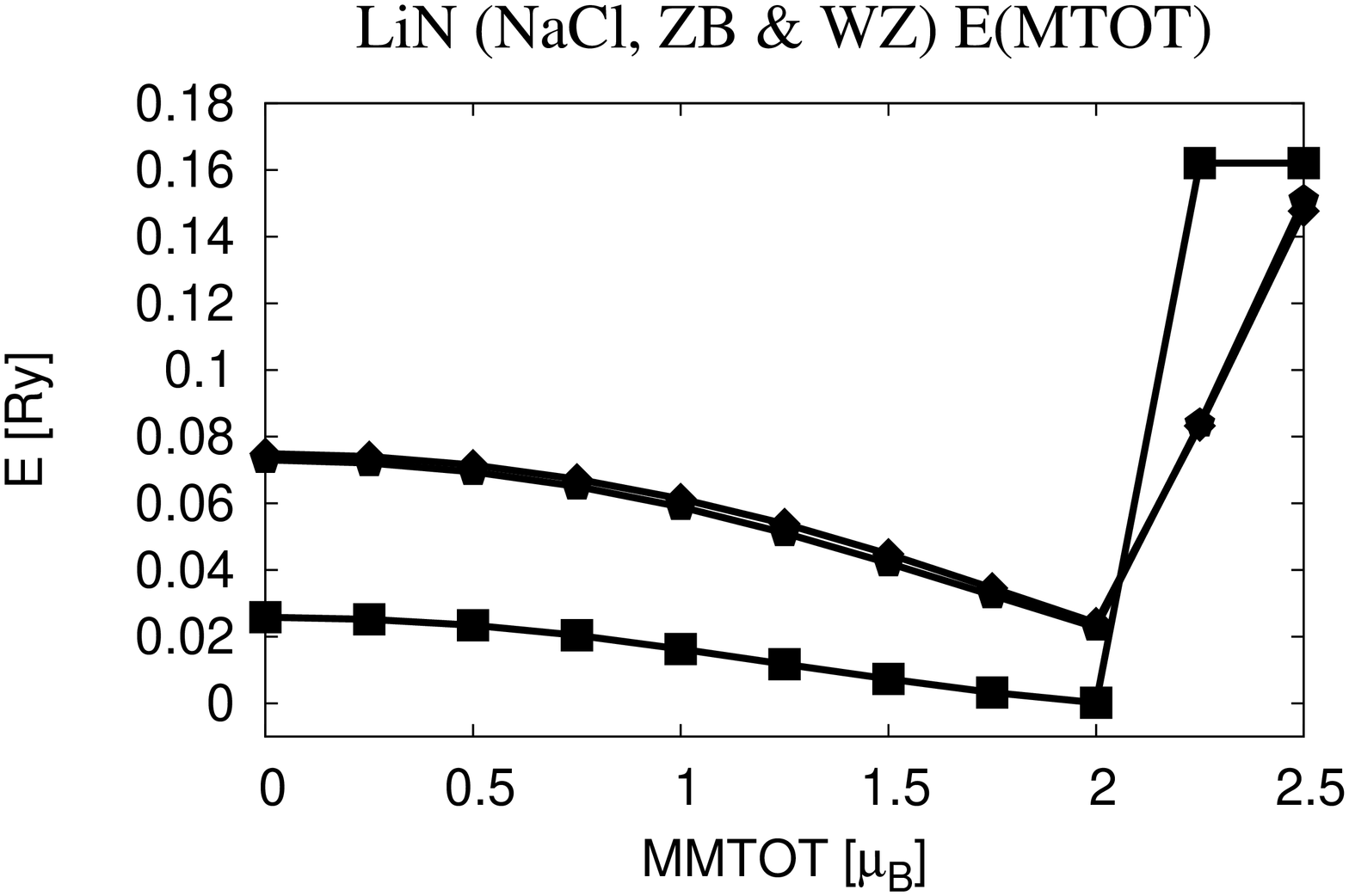}} 
\resizebox{80mm}{!}{\includegraphics[angle=270]{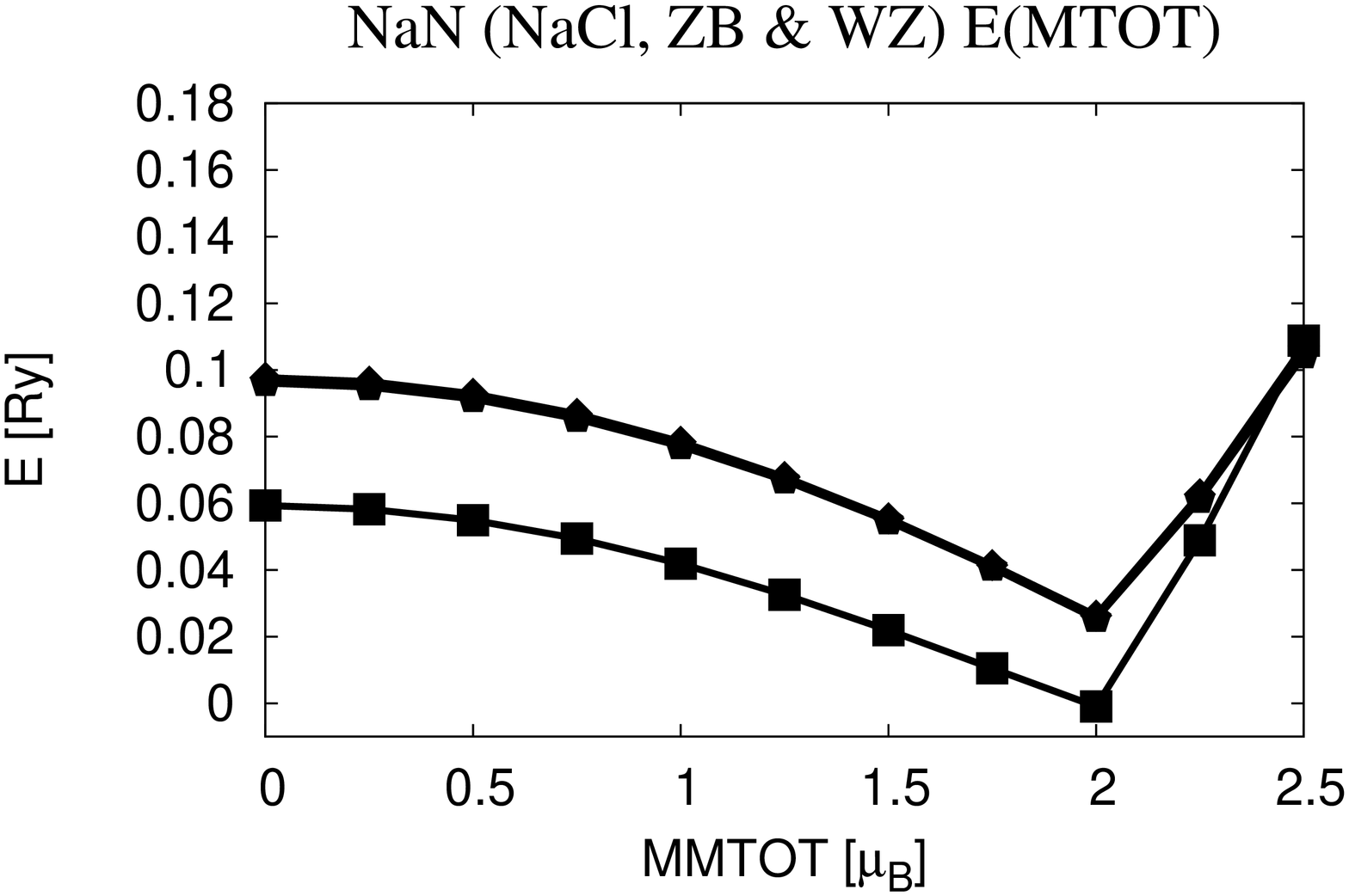}} &  \\ 
\resizebox{80mm}{!}{\includegraphics[angle=270]{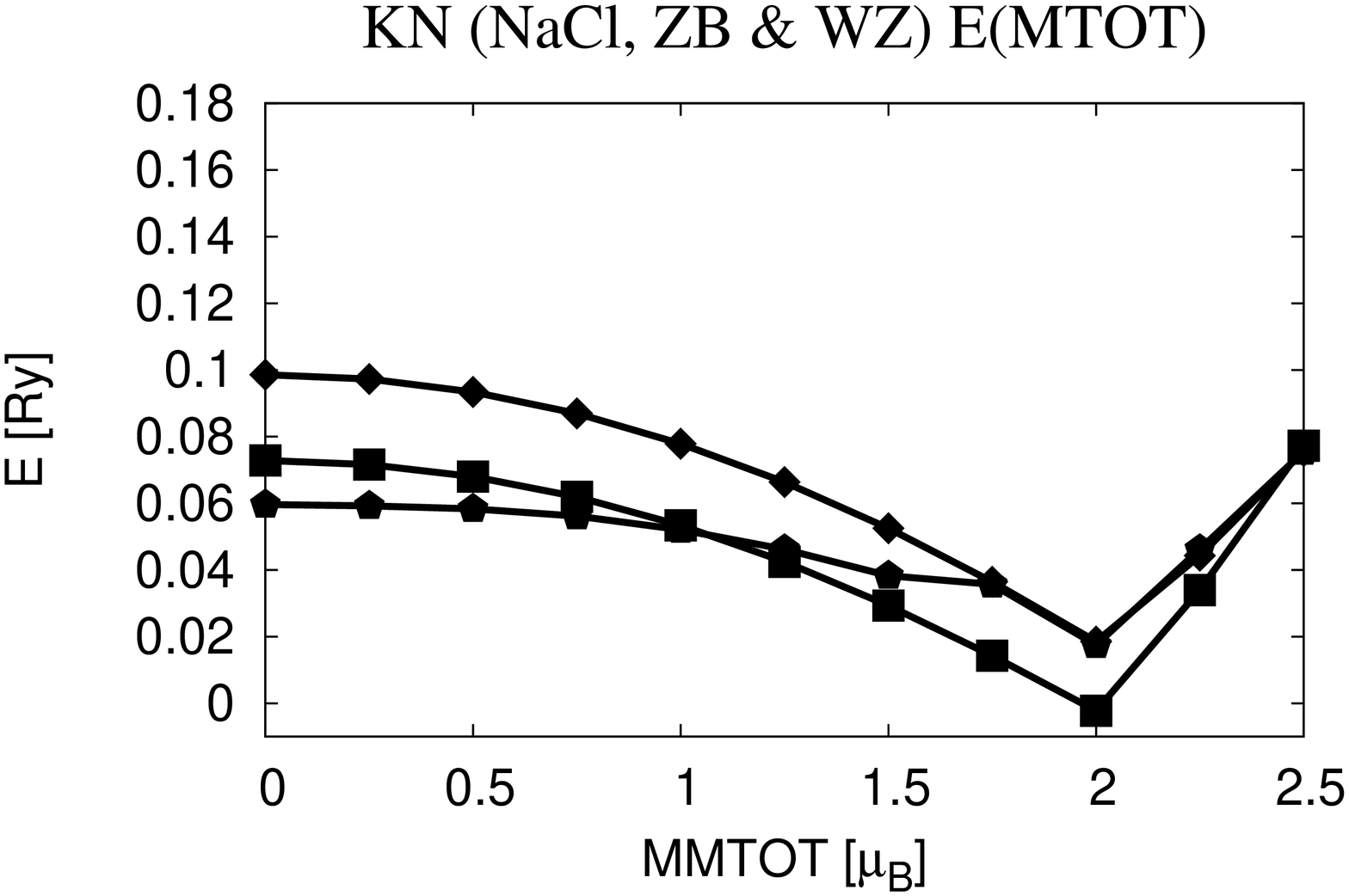}} 
\resizebox{80mm}{!}{\includegraphics[angle=270]{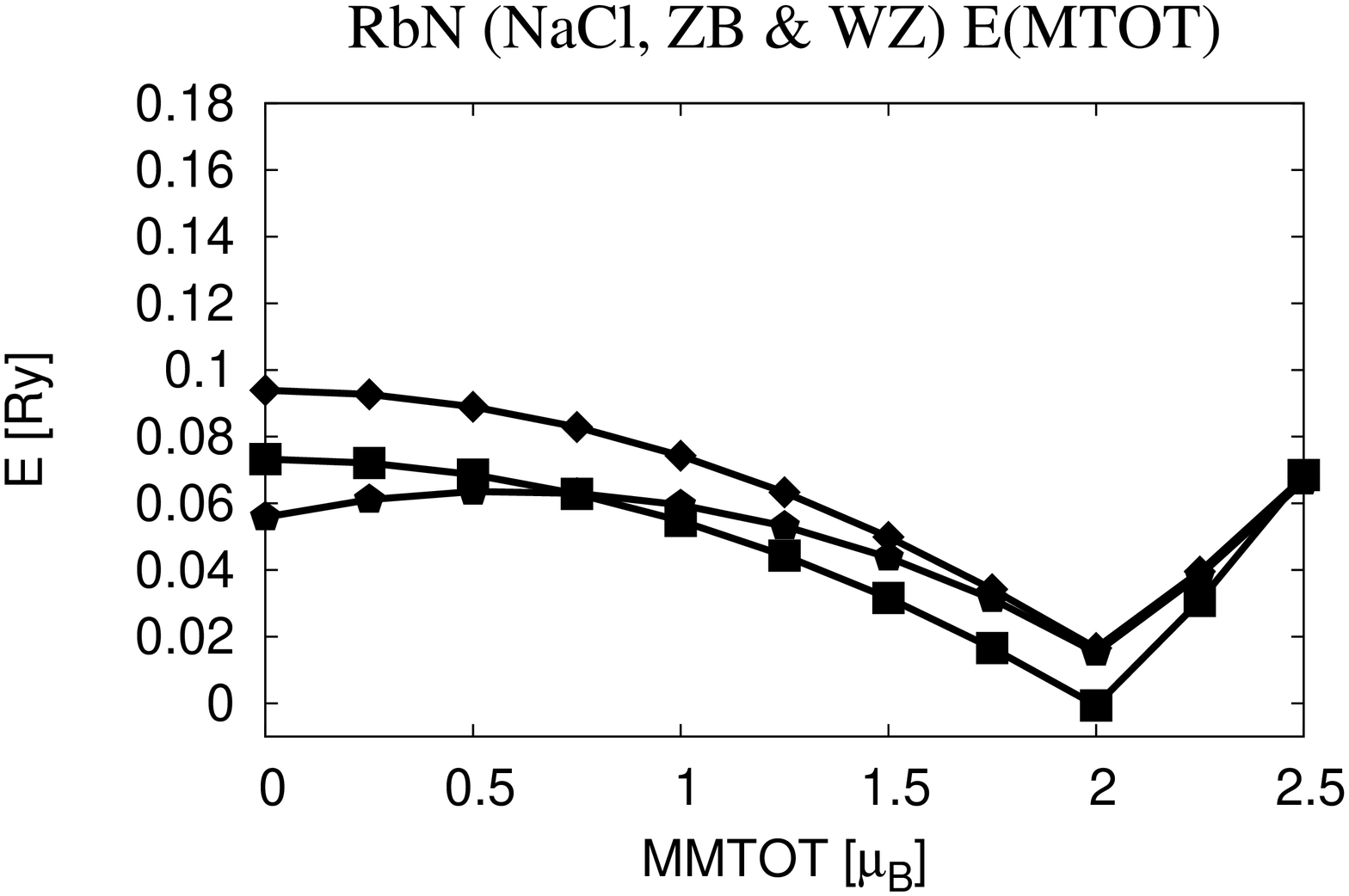}} &  \\ 
& 
\end{tabular}
\end{center}
\caption{Total energy \textit{E} (in Ry), relative to the minimum energy,
for LiN, NaN, KN and RbN in equilibrium for three crystal structures
(squares -- RS, pentagons -- WZ, diamonds -- ZB) as a function of the assumed
net magnetic moment $\protect\mu_{\text{tot}}$ (in $\protect\mu_{\text{B}}$
). Formation of integer magnetic moment $\protect\mu_{\text{tot}}= 2\protect
\mu_{\text{B}}$ stabilizes the crystal structure by increasing chemical
bonds between I$^{A}$ and N atoms.}
\end{figure}

\begin{figure}[H]
\begin{center}
\begin{tabular}{cc}
\resizebox{80mm}{!}{\includegraphics[angle=270]{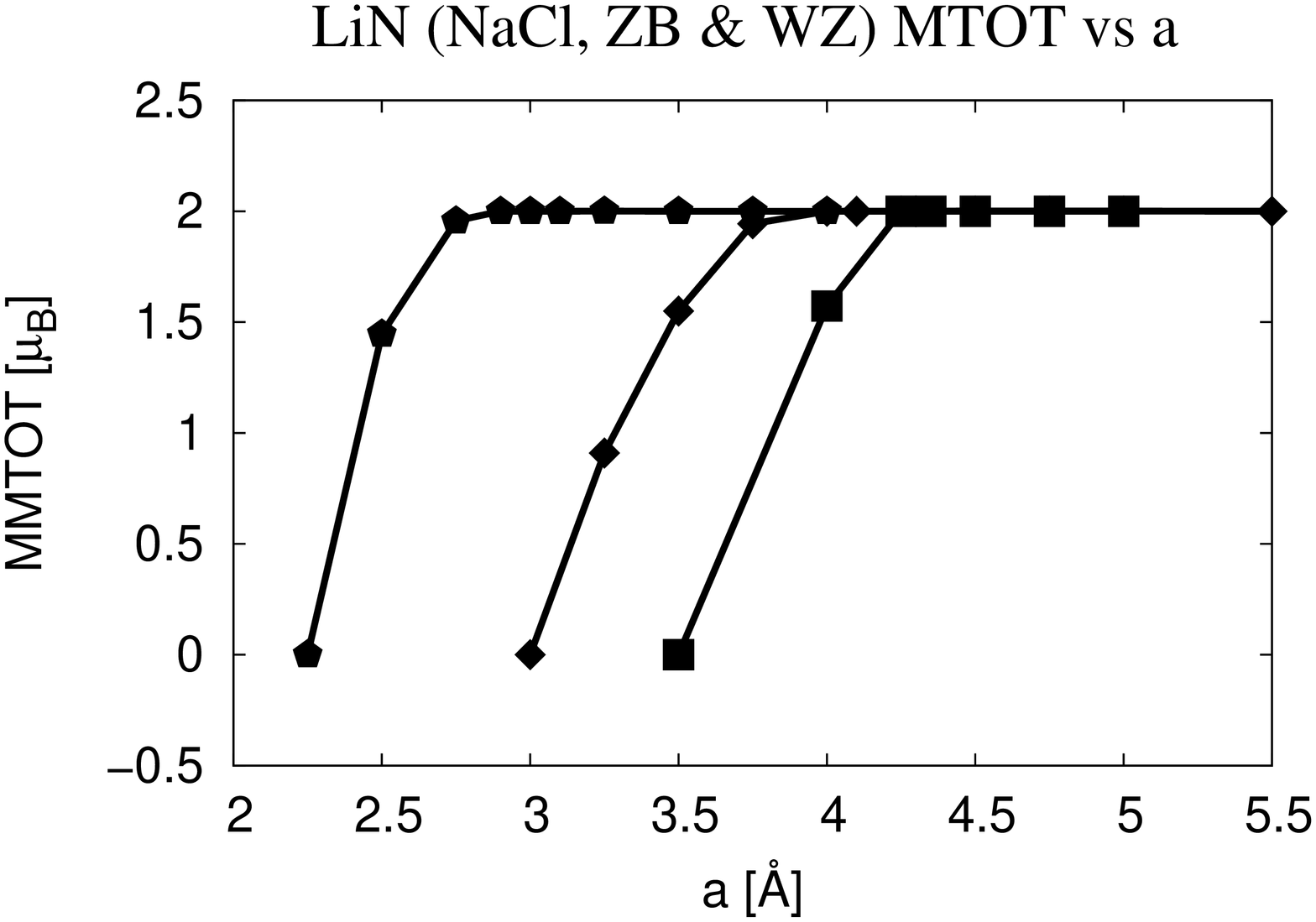}} %
\resizebox{80mm}{!}{\includegraphics[angle=270]{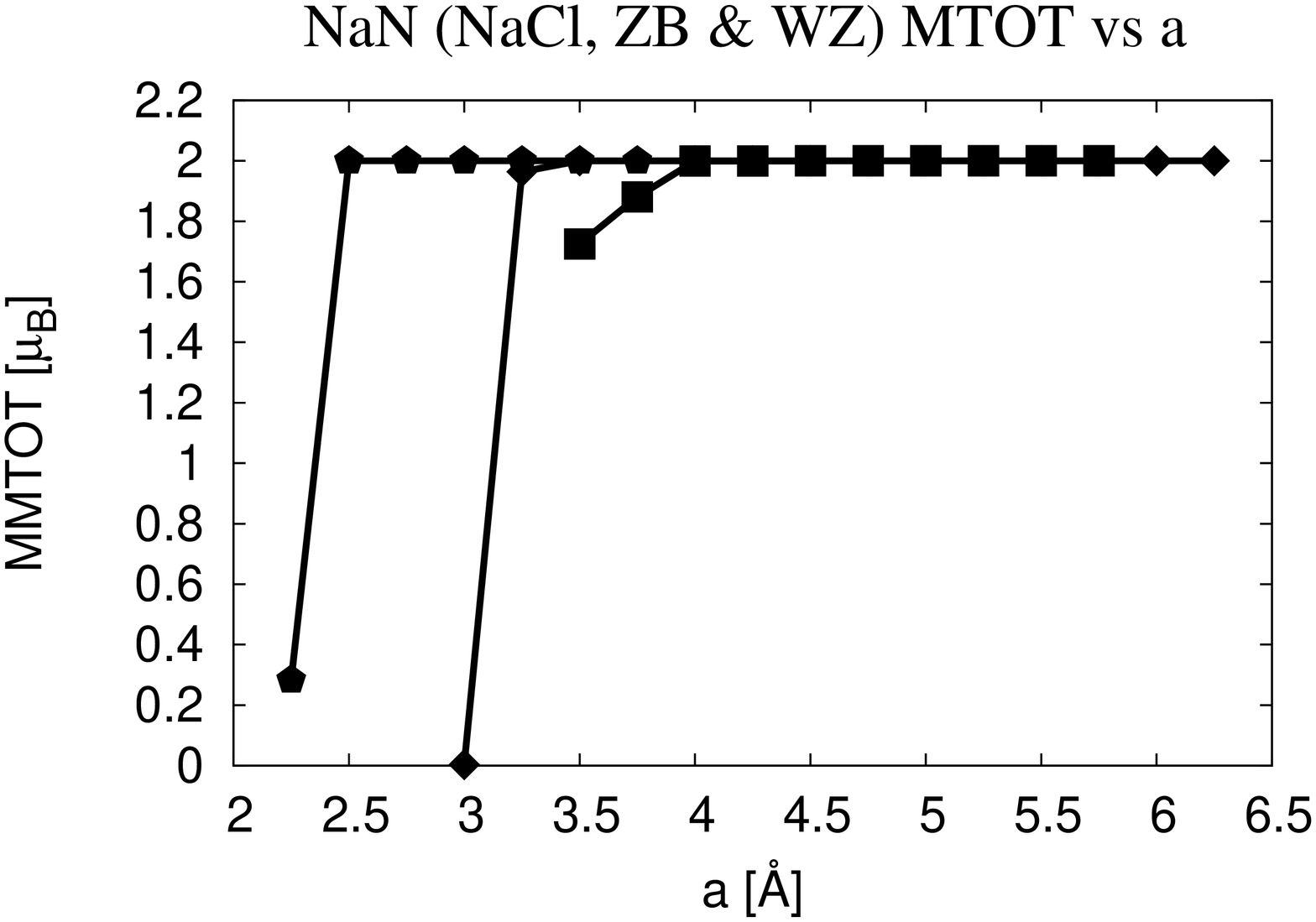}} &  \\ 
\resizebox{80mm}{!}{\includegraphics[angle=270]{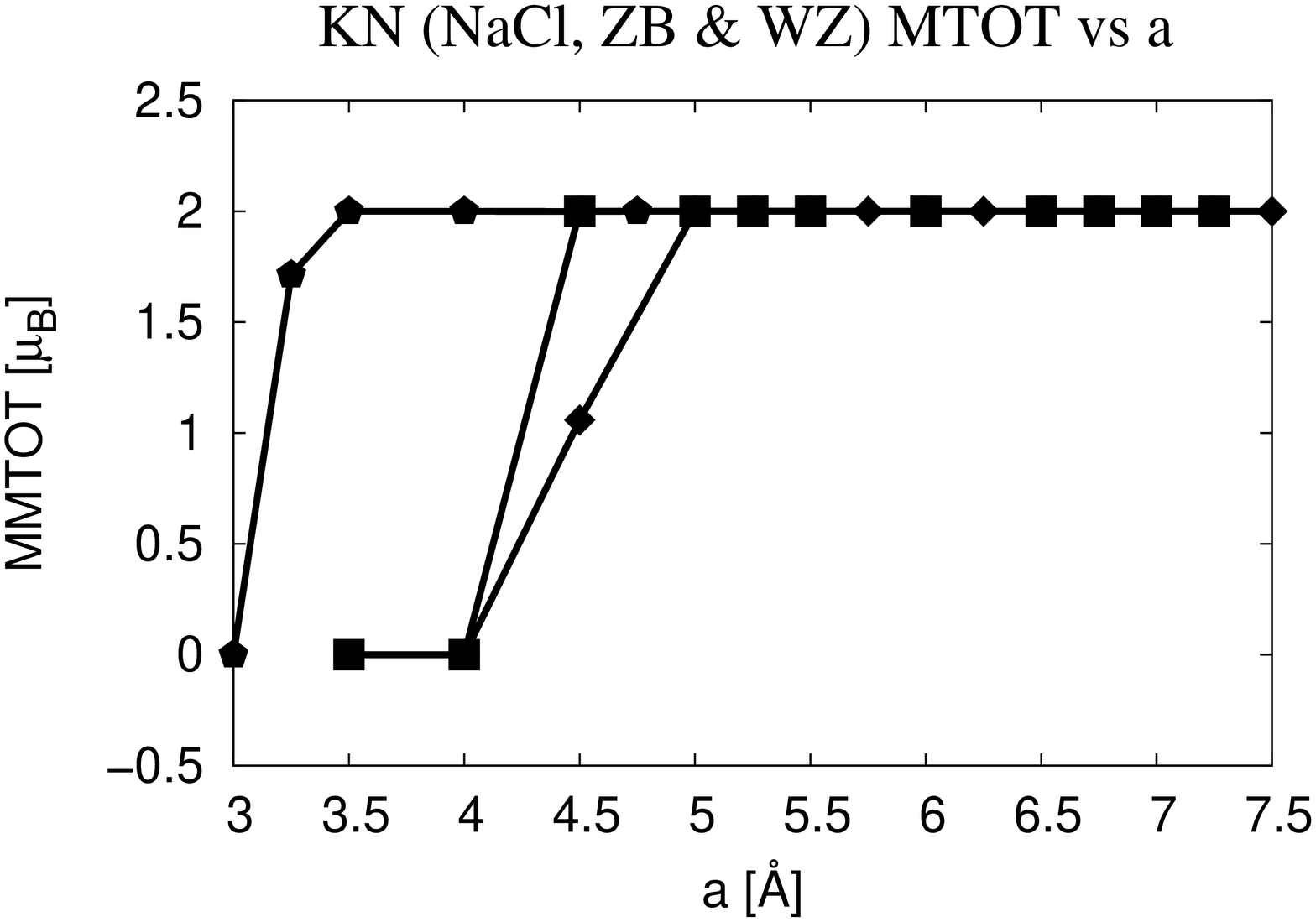}} %
\resizebox{80mm}{!}{\includegraphics[angle=270]{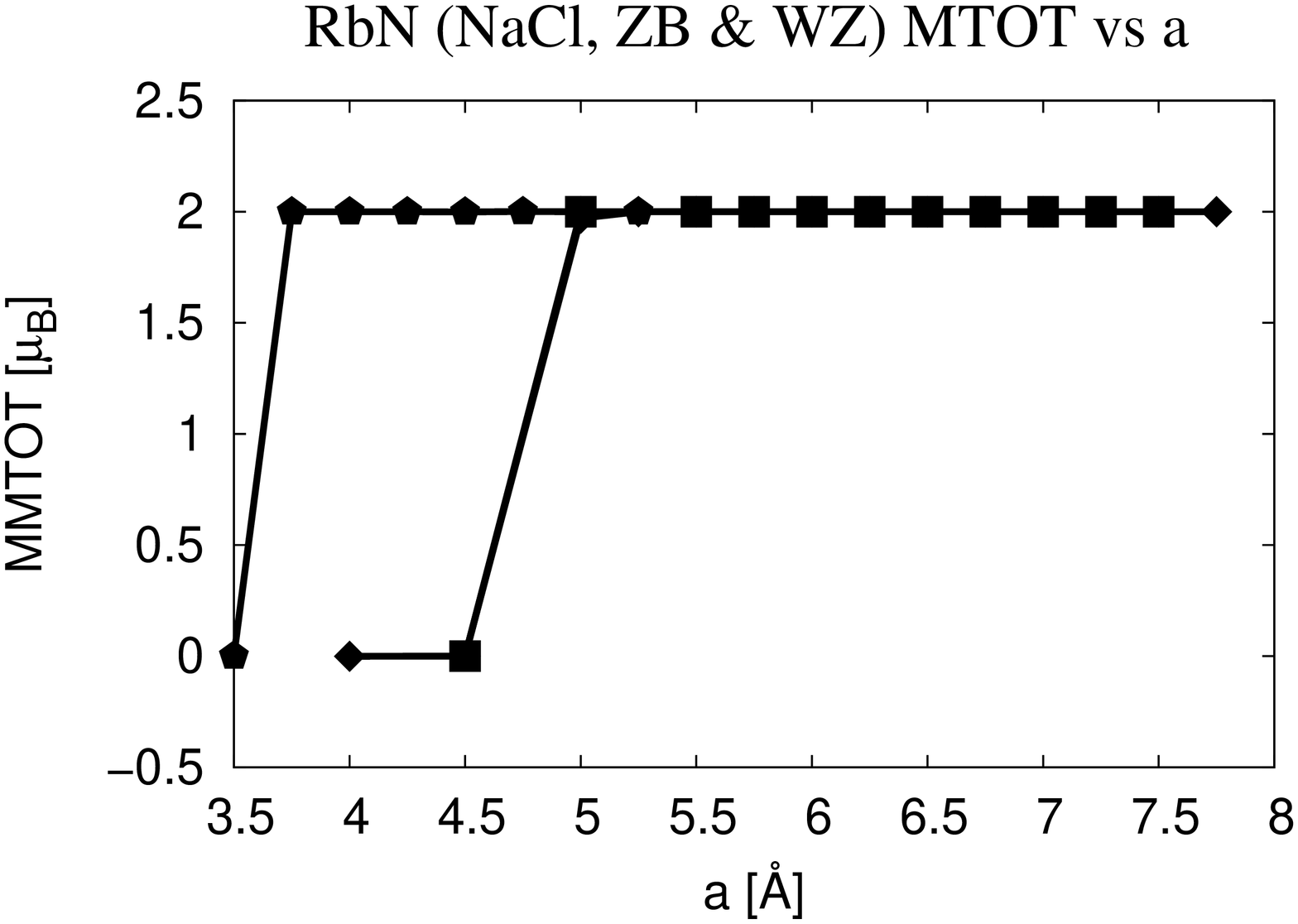}} &  \\ 
& 
\end{tabular}
\end{center}
\caption{Magnetic moment $\protect\mu_{\text{tot}}$ (in $\protect\mu_{\text{B%
}}$) vs. lattice constant \textit{a} (in $\ {\rm\AA}$) for LiN,
NaN, KN and RbN in three crystal structures (squares -- RS, pentagons -- WZ,
diamonds -- ZB). The lines are to guide the eye. Critical values of lattice
parameter \textit{a} for compression, below which the total magnetic moment $
\protect\mu_{\text{tot}}=2\protect\mu_{\text{B}}$ disappears, can be seen
for each compound in three crystal structures. For example, in the case of
LiN the critical values of \textit{a} are 2.75$\ {\rm\AA}$, 3.75
$\ {\rm\AA}$ and 4.25$\ {\rm\AA}$ for WZ, ZB and
RS structures, respectively.}
\end{figure}

\begin{figure}[H]
\begin{center}
\begin{tabular}{cc}
\resizebox{80mm}{!}{\includegraphics[angle=0]{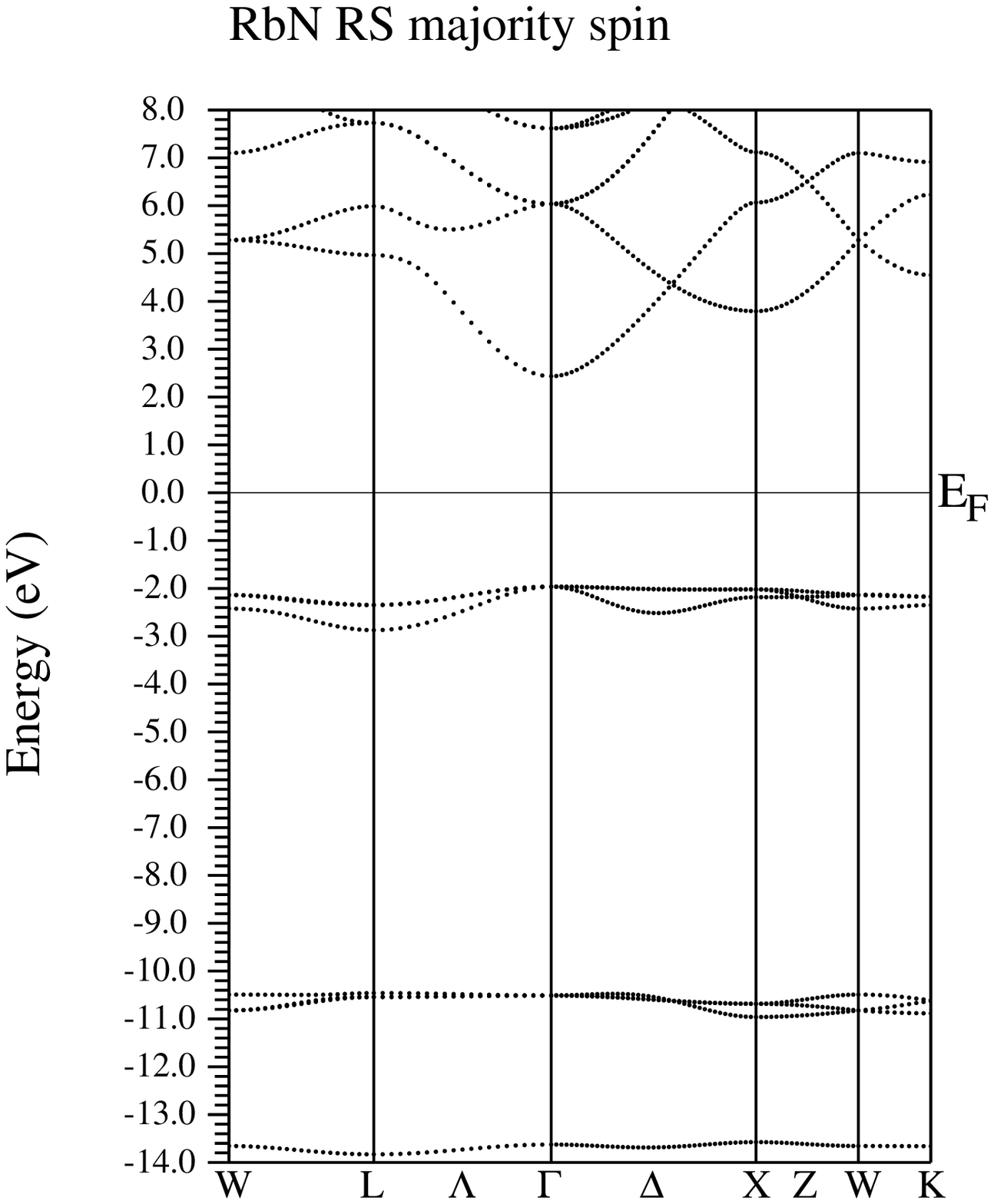}} 
\resizebox{80mm}{!}{\includegraphics[angle=0]{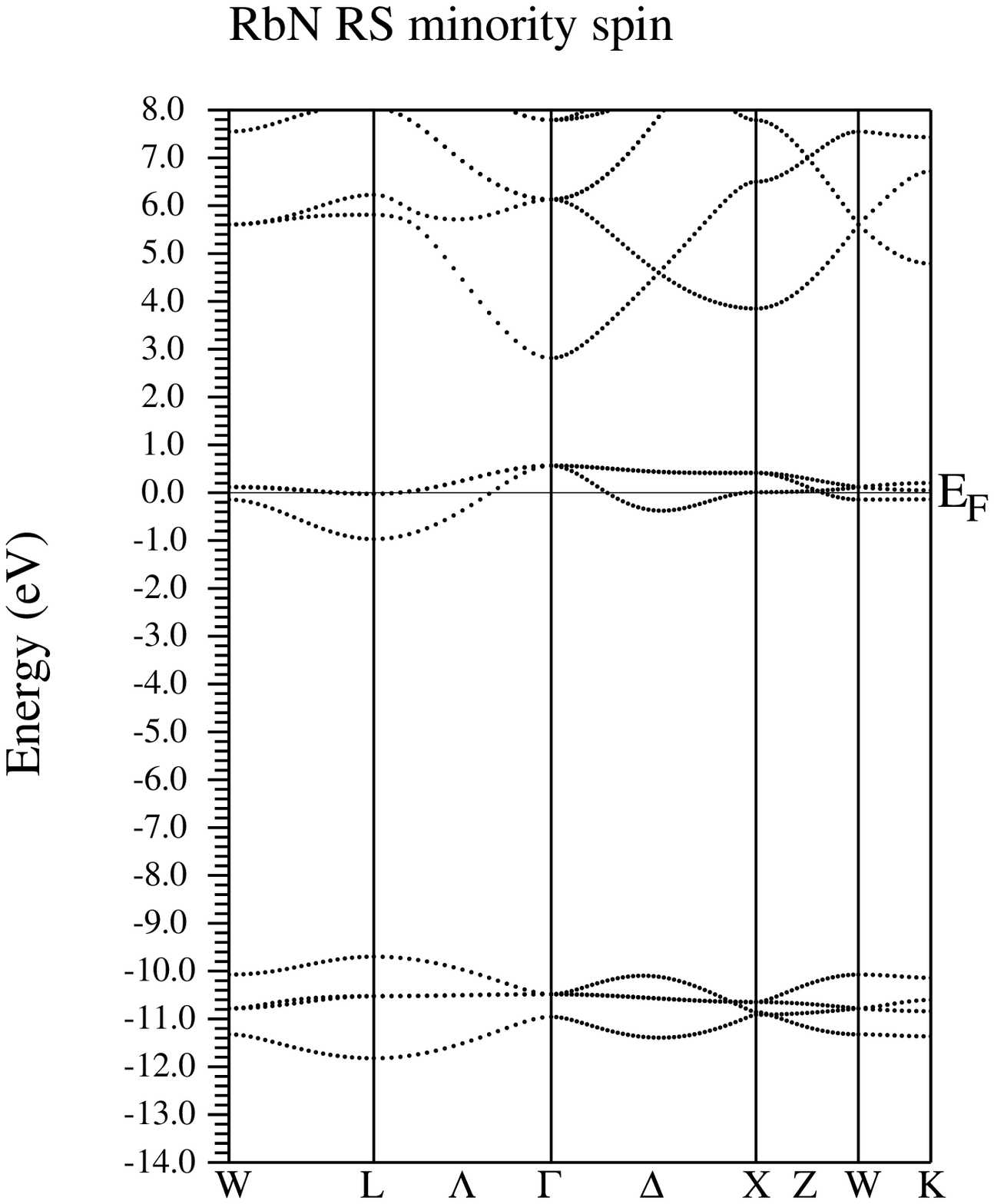}} &  \\ 
\resizebox{80mm}{!}{\includegraphics[angle=0]{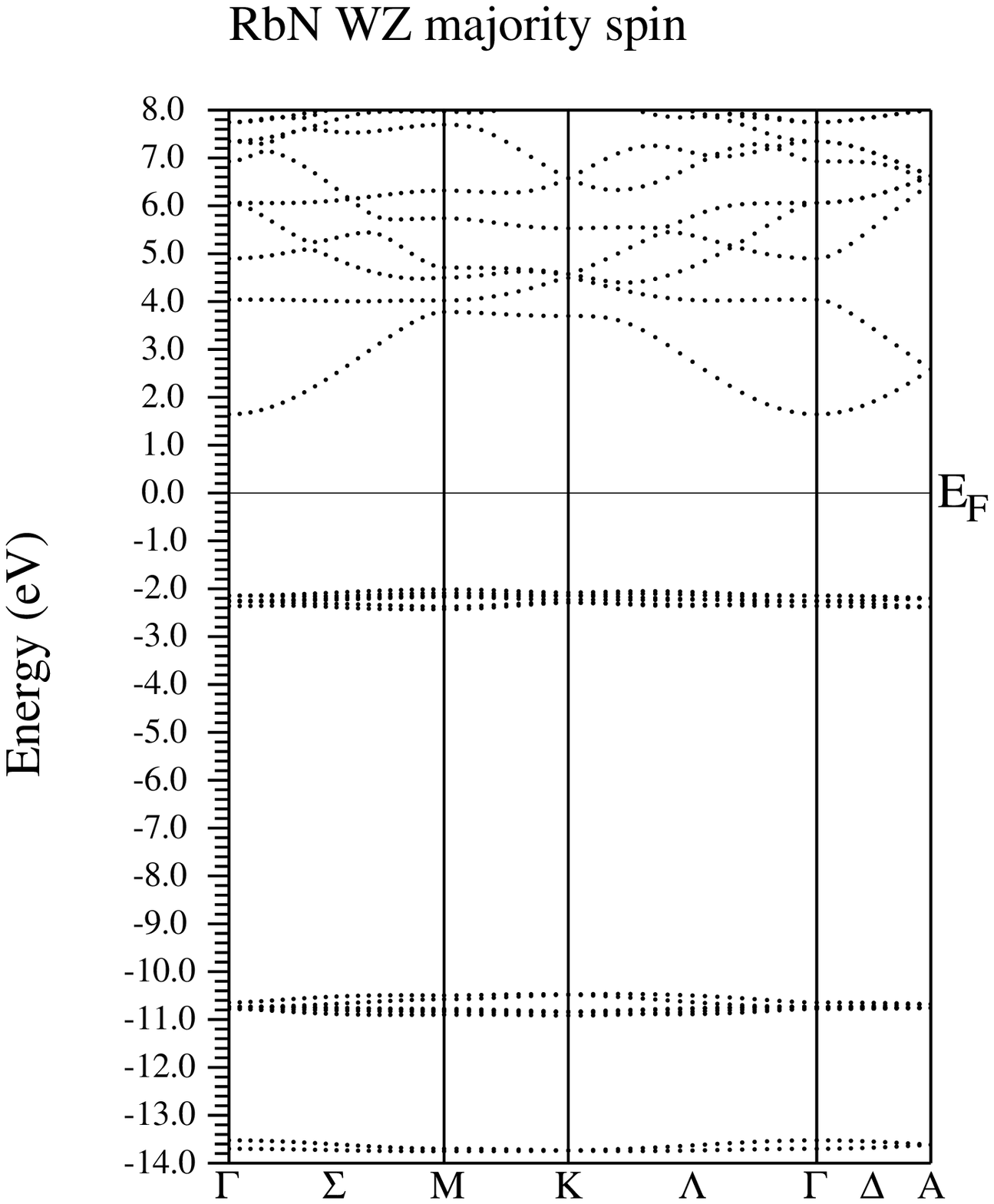}}
\resizebox{80mm}{!}{\includegraphics[angle=0]{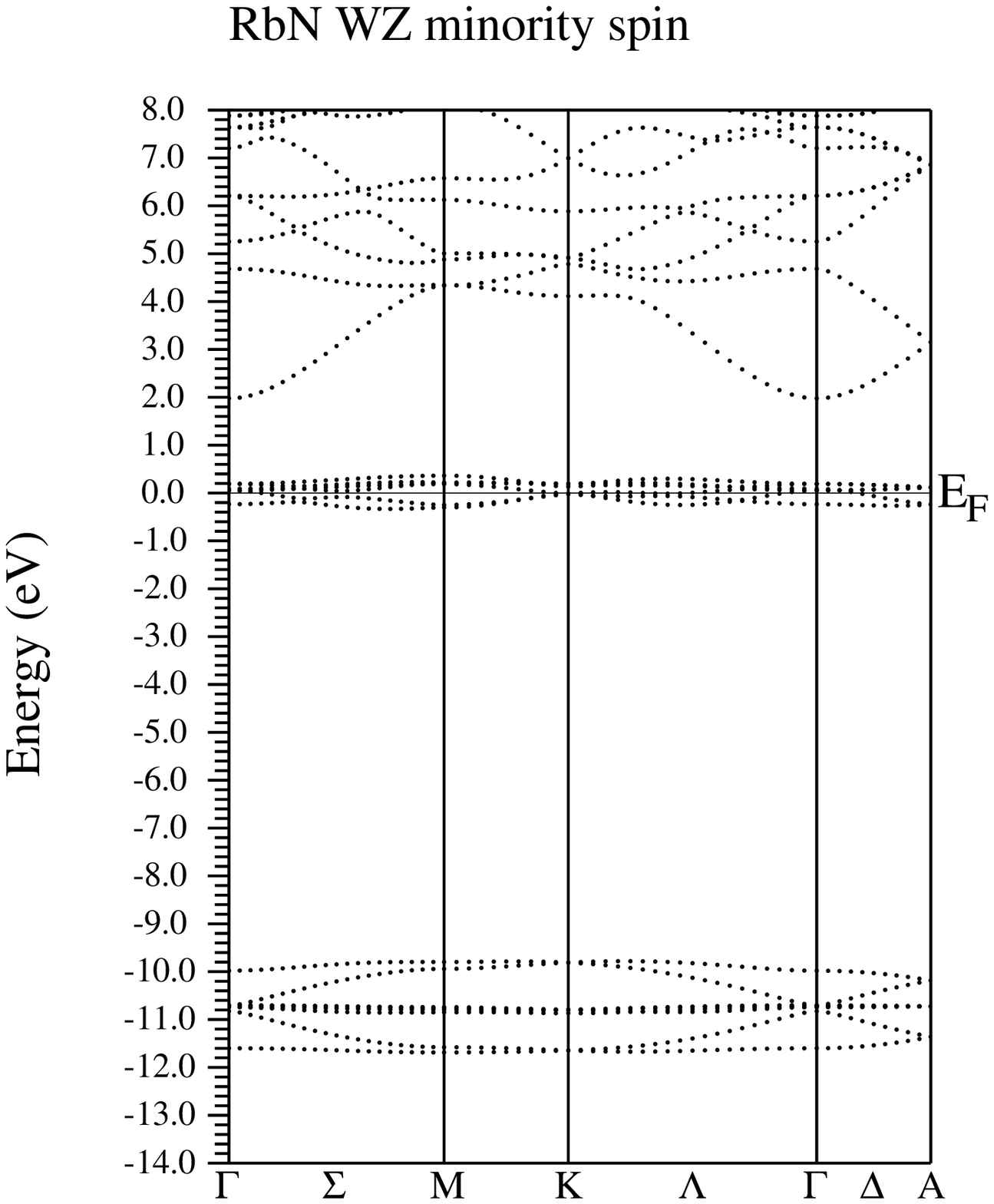}} &  \\ 
& 
\end{tabular}
\caption{Spin resolved electron energy bands for RbN in different crystal structures. High symmetry points from the Brillouin zone for two crystal structures are indicated.
Highest occupied majority spin (up) subbands, shown in the left-hand-side diagrams, are separated from
the Fermi level $E_\text{F}$ by half-metallic bandgap $E_{\text{g}}^{\text{HM%
}}=2$eV. The values of bandgaps separating occupied (or partially occupied)
and empty states for both majority $E_{\text{g}}^{\text{up}}$ and minority
(down) $E_{\text{g}}^{\text{dn}}$ spin subbands (shown in the right-hand
side diagrams) are listed in Table II. Corresponding total and partial DOS
are presented in Figure~6.}
\end{center}
\end{figure}

\begin{figure}[H]
\begin{center}
\begin{tabular}{cc}
\resizebox{80mm}{!}{\includegraphics[angle=0]{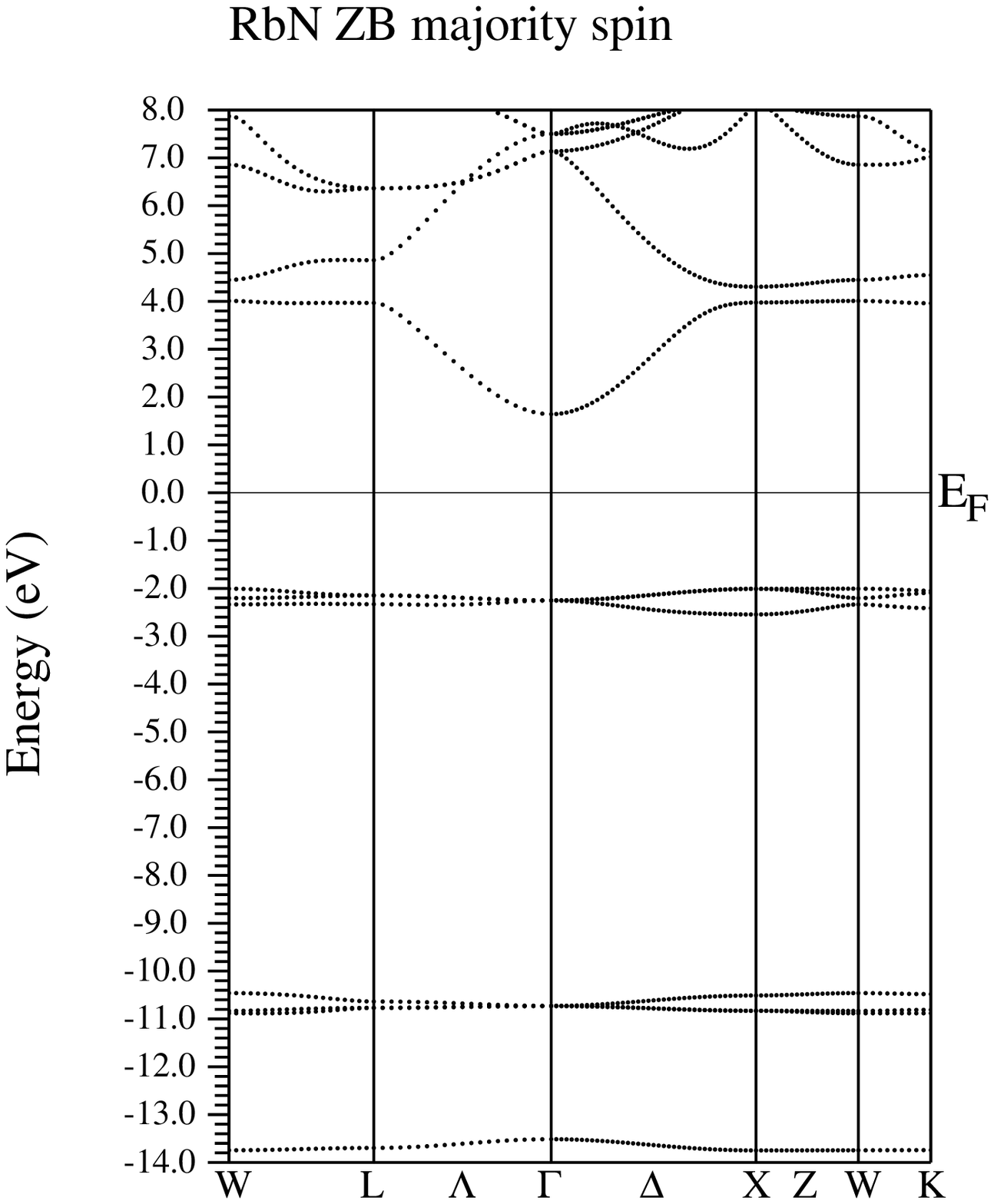}} 
\resizebox{80mm}{!}{\includegraphics[angle=0]{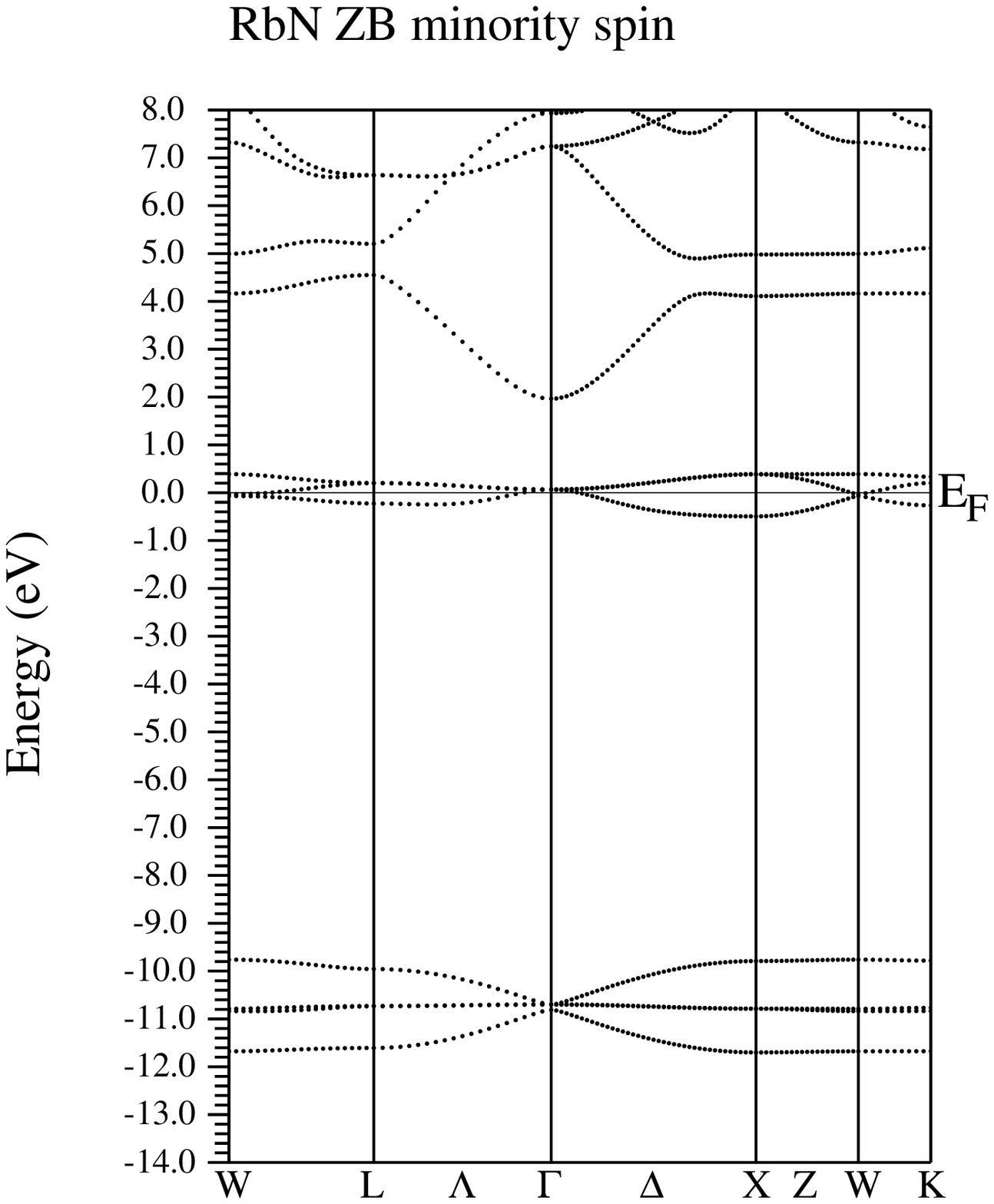}} &  \\ 
\resizebox{80mm}{!}{\includegraphics[angle=0]{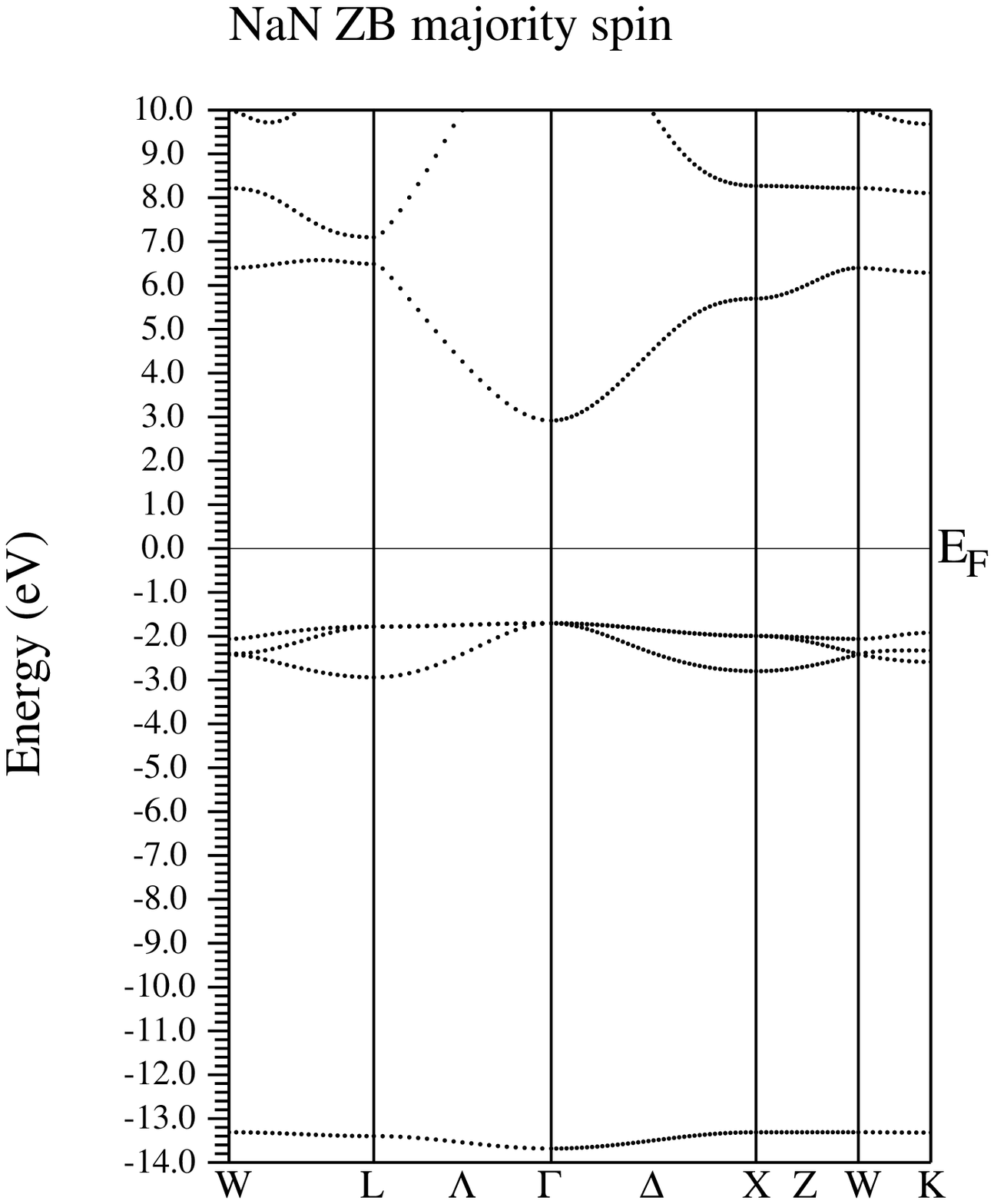}} 
\resizebox{80mm}{!}{\includegraphics[angle=0]{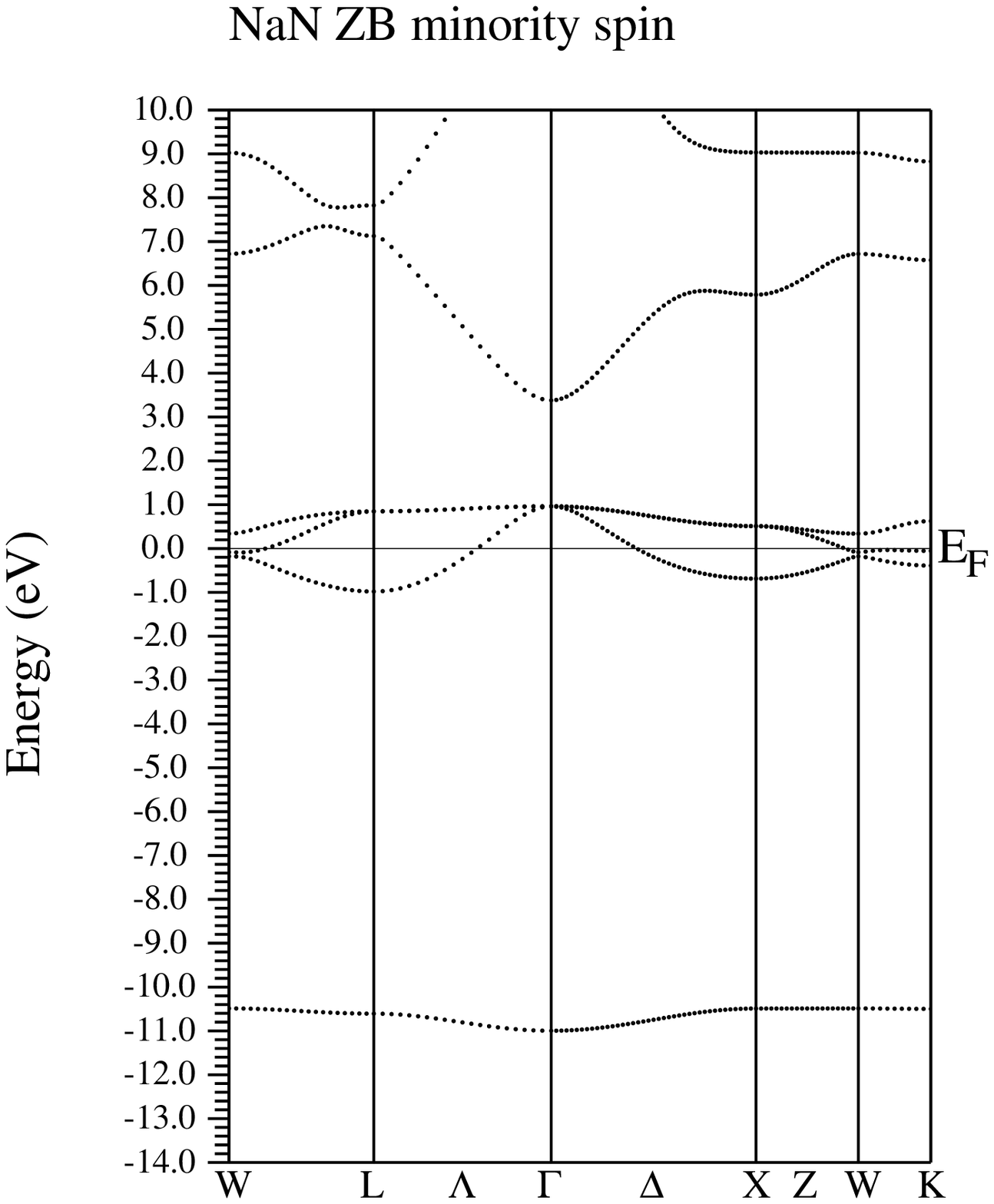}} &  \\ 
& 
\end{tabular}
\caption{Spin resolved electron energy
bands for RbN and NaN in ZB crystal structure. High symmetry points from the Brillouin zone for ZB crystal structure are indicated.
Highest occupied majority
spin (up) subbands, shown in the left-hand-side diagrams, are separated from
the Fermi level $E_\text{F}$ by half-metallic bandgap $E_{\text{g}}^{\text{HM%
}}=2$eV. The values of bandgaps separating occupied (or partially occupied)
and empty states for both majority $E_{\text{g}}^{\text{up}}$ and minority
(down) $E_{\text{g}}^{\text{dn}}$ spin subbands (shown in the right-hand
side diagrams) are listed in Table II. Corresponding total and partial DOS
are presented in Figure 6.}
\end{center}
\end{figure}

\begin{table}[H]
\begin{tabular}{lccc}
\hline\hline
\  & $E_{\text{g}}^{\text{up}}$ (eV) & \ $E_{\text{g}}^{\text{dn}}$ (eV) & \ 
$E_{\text{g}}^{\text{HM}}$ (eV) \\ \hline
LiN (RS) & 8.05 & 6.26 & 0.2 \\ 
LiN (WZ) & 6.56 & 4.53 & 0.9 \\ 
LiN (ZB) & 6.55 & 4.52 & 1.2 \\ \hline
NaN (RS) & 5.17 & 3.12 & 1.0 \\ 
NaN (WZ) & 4.38 & 2.20 & 1.7 \\ 
NaN (ZB) & 4.44 & 2.37 & 1.6 \\ \hline
KN (RS) & 4.54 & 2.37 & 2.0 \\ 
KN (WZ) & 3.81 & 1.72 & 2.0 \\ 
KN (ZB) & 3.77 & 1.77 & 2.0 \\ \hline
RbN (RS) & 4.22 & 2.06 & 2.0 \\ 
RbN (WZ) & 3.48 & 1.42 & 2.0 \\ 
RbN (ZB) & 3.45 & 1.39 & 2.0 \\ \hline
\end{tabular}
\caption{Energy bandgaps for majority up-spin ($E_{\text{g}}^{\text{up}}$)
and minority down-spin ($E_{\text{g}}^{\text{dn}}$) states together with
half-metallic bandgaps ($E_{\text{g}}^{\text{HM}}$) for three crystal
structures.}
\end{table}

\begin{figure}[H]
\begin{center}
\begin{tabular}{cc}
\resizebox{80mm}{!}{\includegraphics[angle=270]{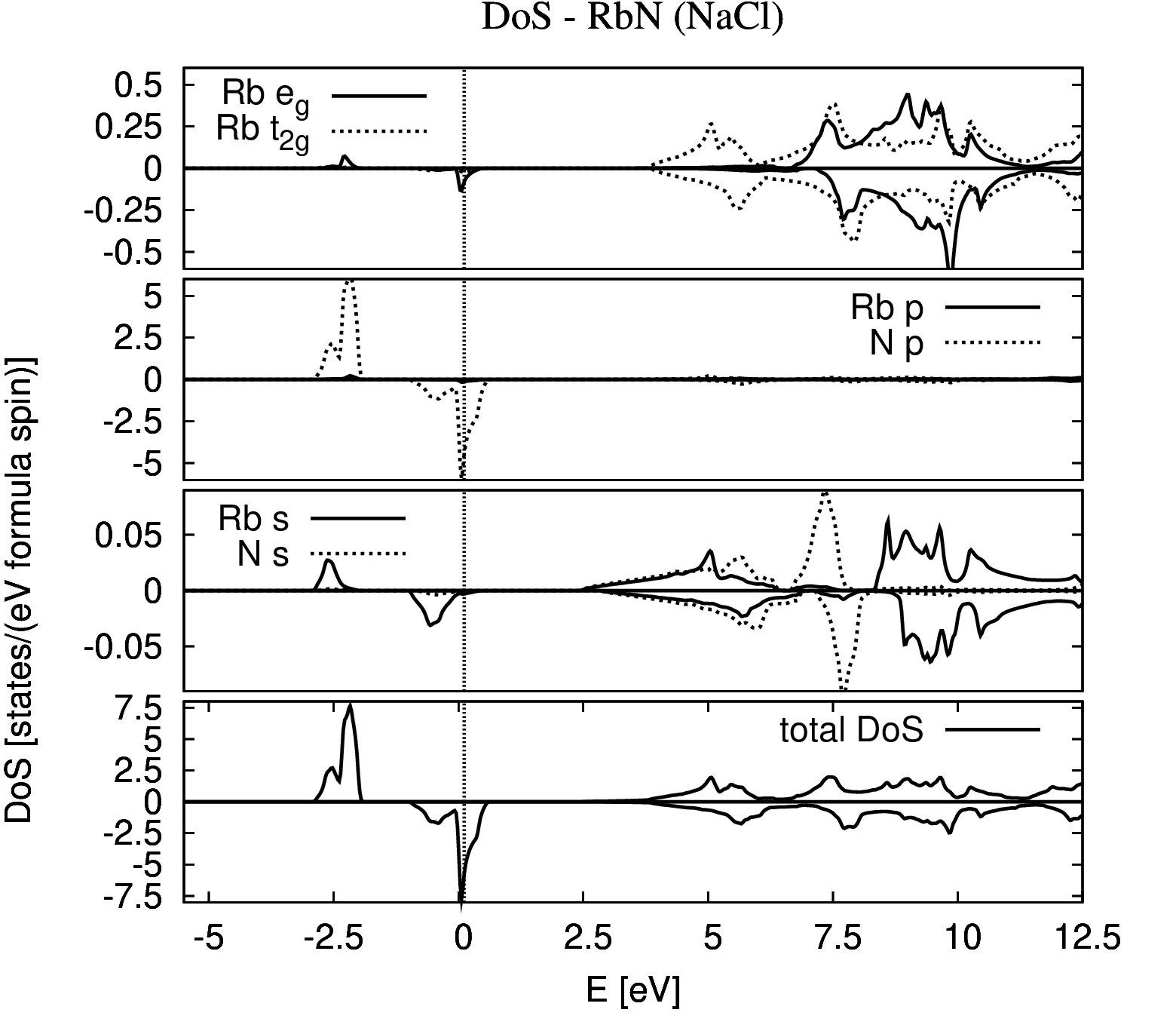}} %
\resizebox{80mm}{!}{\includegraphics[angle=270]{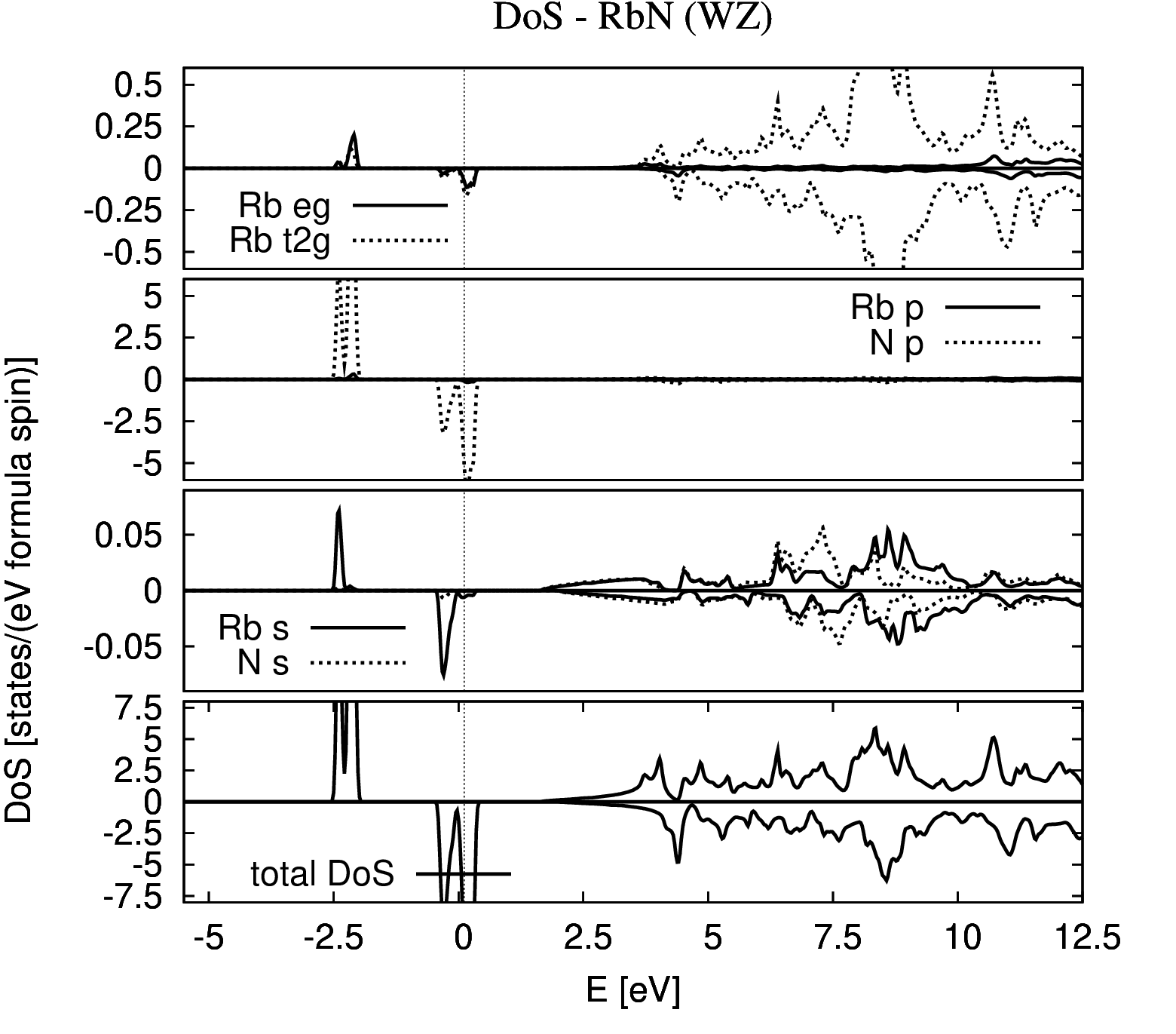}} &  \\ 
\resizebox{80mm}{!}{\includegraphics[angle=270]{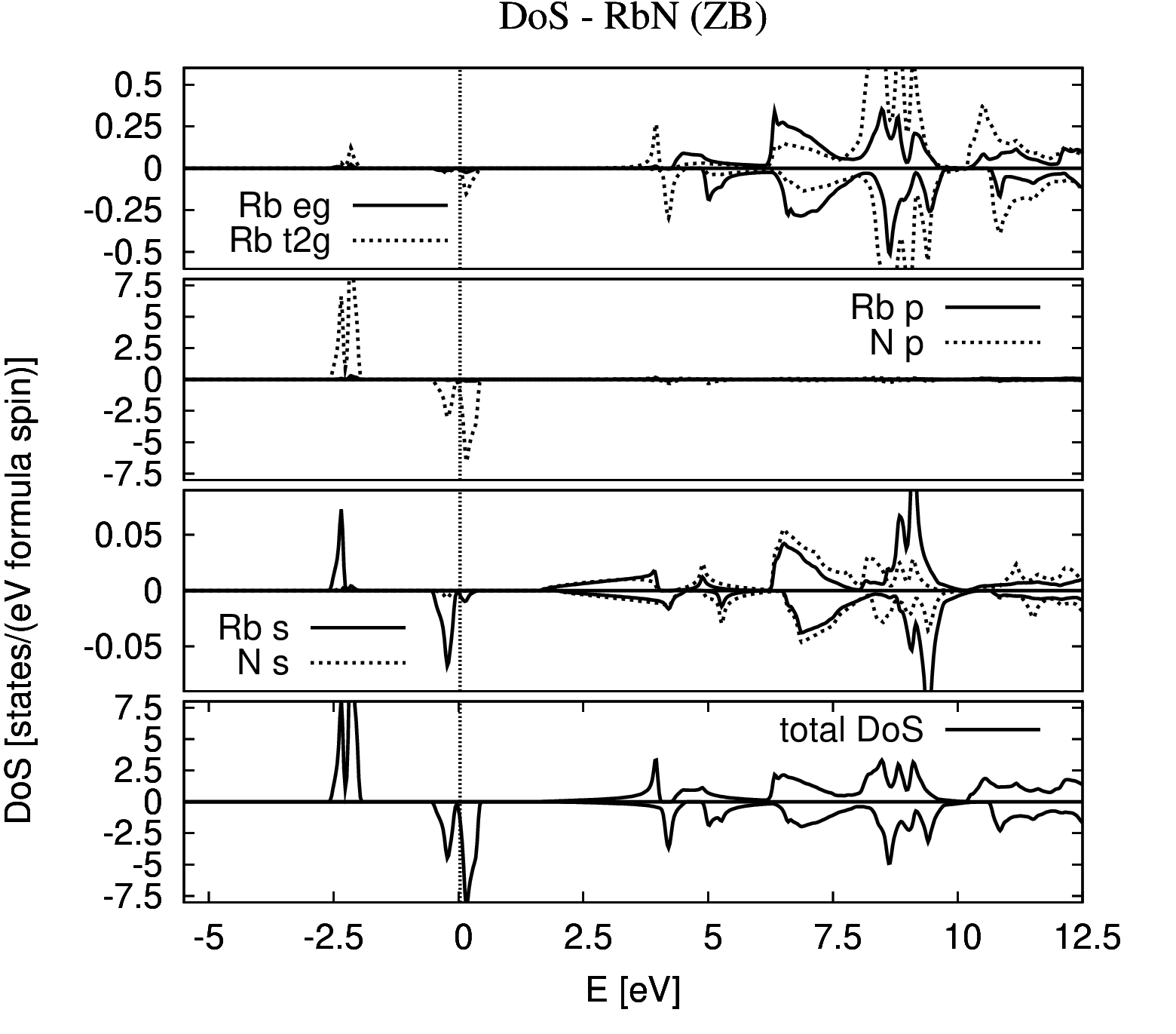}} %
\resizebox{80mm}{!}{\includegraphics[angle=270]{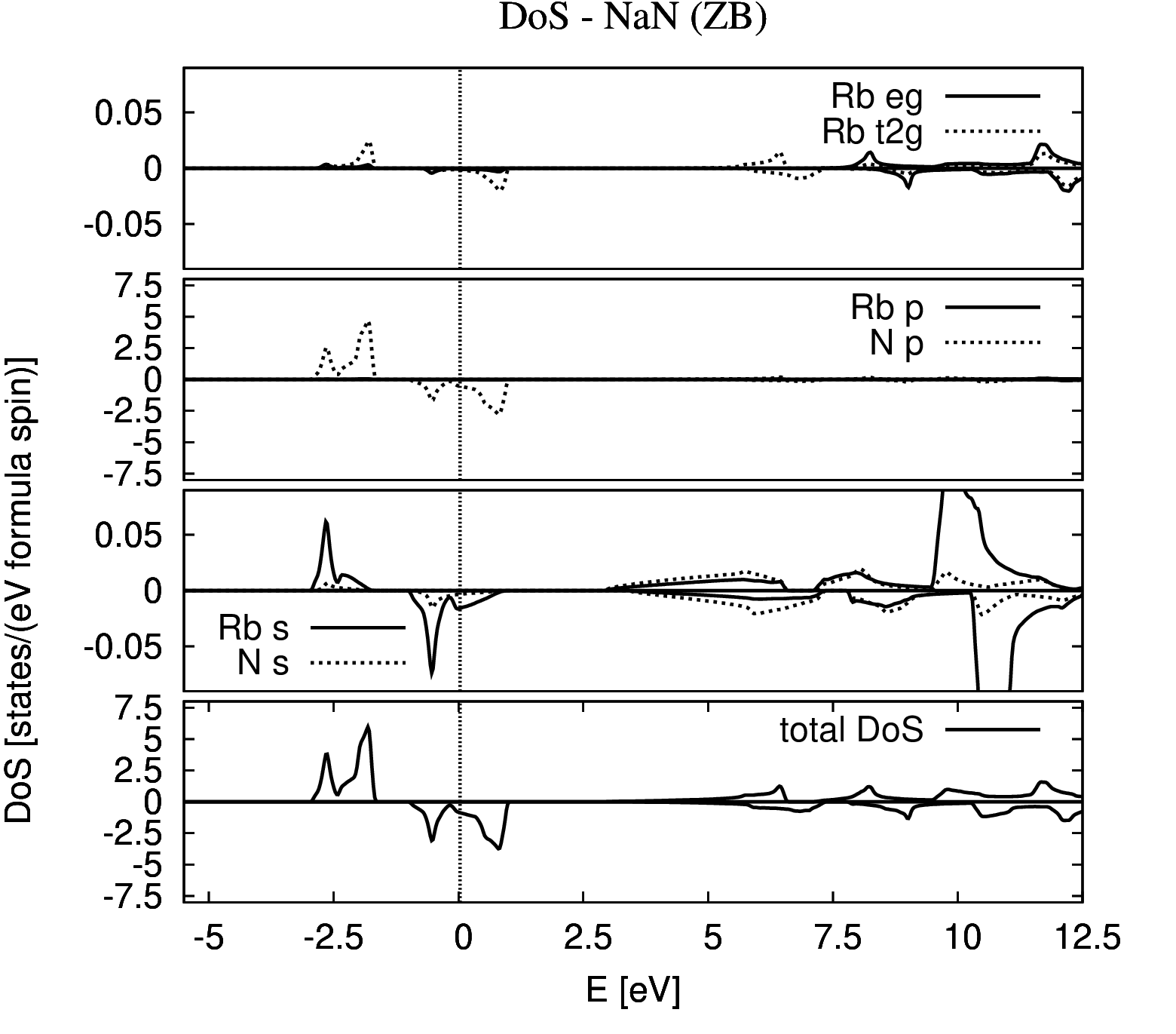}} &  \\ 
& 
\end{tabular}
\end{center}
\caption{Total and partial site and symmetry projected DOS for RbN and NaN in different crystalline
structures. Angular electron density distribution in the real space for the $%
d_{x^{2}-y^{2}}$ state is in $x$ and $y$ axis, for the state $%
d_{3z^{2}-r^{2}}$ in $z$ axis, what increases the overlap with N $p$
electron states in the RS structure and explains the hybridization between N 
$p$ and $e_{g}$ states. On the other hand, angular dependence of $t_{2g}$
states is favored in the ZB structure because of tetrahedral surrounding of
neighboring atoms and supports hybridization between N $p$ and $t_{2g}$
states. The hexagonal z axis in WZ structure distinguishes both $t_{2g}$ and 
$e_{g}$ Rb states as partners for hybridization with N $p$ states.}
\end{figure}

\section{Density of states and HM ferromagnetism}

It is instructive to compare the band structure (Figs. 4 and 5) with spin resolved total and partial site and symmetry projected density of states (DOS) presented in Figure 6. There are six valence electrons in RbN (Rb: $5s^{1}$; N: $2s^{2}2p^{3}$). Two of them occupy the low-energy N $s$ states, about 11 eV below the Fermi level, and the remaining four electrons are mainly involved in filling anion N $p$ states. The crystal field of cubic or hexagonal symmetry caused by surrounding N anions splits the Rb cation 4${d}$ states with three $t_{2g}$ states ($d_{xy}, d_{yz}, d_{zx}$) lying lover in energy and the two $e_{g}$ states ($d_{x^{2}-y^{2}}, d_{3z^{2}-r^{2}}$) lying higher. Hybridization between N \textit{p} states
and Rb $t_{2g}$ or $e_{g}$ 4\textit{d} states (depending
on the type of crystal lattice) creates both bonding and antibonding hybrid
orbitals. The bonding orbital, lower in energy, appears at the edge of
normally occupied region of N \textit{p} states. The antibonding
hybrid orbital remains in the Rb $t_{2g}$ or $e_{g}$ manifold but
pushed up in energy relative to the nonbonding state. 

The bands around Fermi level are formed mainly of N \textit{p} states with small
admixture of Rb ${d}$ states with predominant $e_{g}$ symmetry in RS structure
and $t_{2g}$ symmetry for ZB and WZ structures for both spin orientations.
In contrast to the \textit{p} states, the $s$ states are placed well below or high above Fermi level giving no net spin polarization. Similarly, the bands located at higher energies (from 4.5 to 12.5 eV above the Fermi level) are mainly composed of \textit{d} states belonging to $t_{2g}$ and $e_{g}$ representations. Hence, the \textit{s} and \textit{d} states are  not directly involved in creation of spin polarisation and ferromagnetism in RbN and other I$^{A}$-N binary compounds.

Existence of narrow bands near Fermi level translates into peaks in the DOS
diagrams. According to qualitative Stoner criterion the high density of
electron states at the Fermi level should stabilize the ferromagnetic order.
Pressure induced lattice contraction produces changes in the band
structure, increase of band widths with increasing kinetic energy,
decrease of the spin splitting and destroying ferromagnetism. 

In the presented here results of the ground state total energy calculations we can see the influence of Coulomb intra-atomic electron repulsion. It produces energy shift $\Delta = 2\ \text{eV}$ between narrow majority and minority spin energy band states, visible as peaks at -- and  below Fermi level in the DOS diagrams, shown in Figure 6. This type of band structure with well separated up and down spin states can be attributed to the Hubbard-like interaction $U=\Delta/2$. The parameter $U=1\ \text{eV}$ is a measure of energy formation for magnetic moment localized at nitrogen ion. According to Hund's rule and Pauli principle
the limiting value of magnetic moment for $p$ electrons is the value of 3 $\mu _{\text{B}}$. The four valence $p$ electrons, present in the system, can produce magnetic moment not greater than  
$(3-1)~\mu _{\text{B}}=2~\mu _{\text{B}}$, what agrees with our calculations. 

Estimation of thermodynamic stability of HM ferromagnetism requires additional calculations to determine the strength of inter-atomic exchange coupling \cite{kubler}. This task requires supercell calculations for planar antiferromagnetic structure of the first (AF1) and second kind (AF2) as particular cases of spiral structure with the \textbf{q} vector $[001]$ and $\frac{1}{2}[111]$, respectively \cite{hmf2}. We have performed relevant calculations only for LiN in WZ crystal structure. The energy differences are $E_{\text{AFM1}}-E_{\text{FM}} = 0.247$ eV and $E_{\text{PARA}}-E_{\text{FM}} = 0.701$ eV. The estimated lover limit for paramagnetic Curie temperature is about 464 K, what makes the material promising for experimental investigation.

\section{Conclusions}

On the basis of \textit{ab initio} calculations employing density functional
theory we have investigated half-metallic ferromagnetism in rock salt,
wurtzite and zinc-blende compounds composed of group I$^{A}$
alkali metals as cations and nitrogen as anion. We
find that, due to the spin polarized \textit{p} orbitals of N, all four
compounds are half-metallic ferromagnets with wide energy bandgaps (up to
2.0 eV). The calculated total magnetic moment in all investigated compounds
is exactly 2.00 $\mu _{\text{B}}$ per formula unit. 
Our calculations show that the predicted half-metallicity is robust
with respect to lattice constant contraction. The formation
of ferromagnetic order requires large lattice constants, high ionicity, empty
$d$ orbitals and slight hybridization between N anion $p$ states and I$^{A}$
cation $d$ states with energies in vicinity of the Fermi level. It is interesting to note
that palladium (with its eight 4$d$ electrons), as dopant replacing Ga atom in WZ GaN semiconductor, interacts with surrounding N atoms in a similar way like alkali atoms do, but 
hybridisation between Pd 4${d}$ and N $p$ orbitals leads to formation of ferromagnetic order with Pd as the main contributor (not N) to the total magnetic moment \cite{osuch}. 

Demonstrated here ferromagnetic order is always more energetically 
stable than antiferromagnetic and paramagnetic state, what makes these materials possible
candidates for spin injection in spintronic devices. Calculations of the
total energy indicate that this class of materials can exist in stable or metastable phase. Their highly interesting magnetic properties should encourage experimentalists to stabilize these materials in properly coordinated structures via vacuum molecular beam epitaxy, chemical transport or laser deposition on suitable substrates.


\newpage 

\end{document}